\documentclass[journal=chreay,manuscript=article]{achemso}

\usepackage[version=3]{mhchem} 
\usepackage{amsmath}
\usepackage{amssymb}
\usepackage{xcolor} 
\usepackage[normalem]{ulem} 

\newcommand*\bi{\begin{itemize}}
\newcommand*\ei{\end{itemize}}
\newcommand*{\BUDGET}{\omega}

\def\md{\mathrm d}
\def\Fo{\text{F}_{\text{o}}}
\def\F1{\text{F}_1}
\def\kT{k_{\rm B}T}
\def\mcP{\mathcal P}
\def\affin{\BUDGET_{\rm tot}}
\def\affinMax{\affin^{\text{max}}}
\newcommand{\fix}{}
\newcommand{\stkout}[1]{}

\author{Aidan I.~Brown}
\email{aibrown@ucsd.edu}
\affiliation{Dept.~of Physics, University of California, San Diego, La Jolla, California, USA}
\author{David A.~Sivak}
\email{dsivak@sfu.ca}
\affiliation{Dept.~of Physics, Simon Fraser University, Burnaby, British Columbia, Canada}

\title[Theory of nonequilibrium free energy transduction by molecular machines]
{Theory of nonequilibrium free energy transduction by molecular machines}

\begin{document}

\begin{tocentry}
\begin{center}
\includegraphics[height=3.5cm]{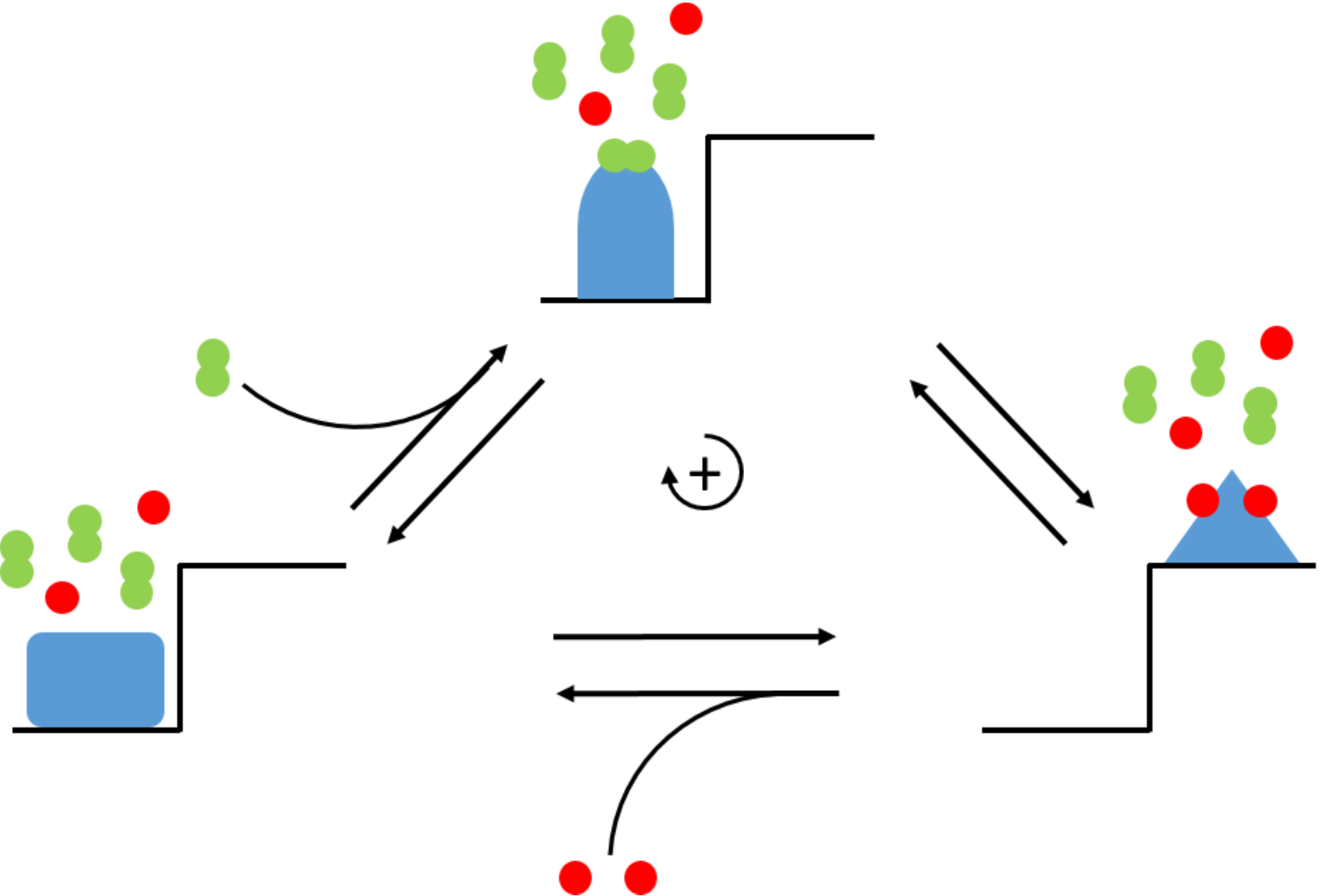}
\end{center}

\end{tocentry}

\begin{abstract}
Biomolecular machines are protein complexes that convert between different forms of free energy. They are utilized in nature to accomplish many cellular tasks. As isothermal nonequilibrium stochastic objects at low Reynolds number, they face a distinct set of challenges compared to more familiar human-engineered macroscopic machines. Here we review central questions in their performance as free energy transducers, outline theoretical and modeling approaches to understand these questions, identify both physical limits on their operational characteristics and design principles for improving performance, and discuss emerging areas of research. 
\end{abstract}

\tableofcontents


\section{Introduction}
A hallmark of all living things is order, 
manifested both in structure and dynamical processes. Such order is not possible at equilibrium, where the second law of thermodynamics requires a maximum of disorder, so living things must fundamentally be out of equilibrium~\cite{schrodinger44}. 
Specifically, cells are characterized by out-of-equilibrium chemical concentrations and 
inhomogeneous
spatial distributions of charge and molecular species~\cite{philips12,alberts14}.

These nonequilibrium conditions are largely created and maintained by molecular machines, macromolecular complexes that interconvert or transduce different stores (reservoirs) of nonequilibrium free energy.
Molecular machines play essential roles in a panoply of core cell-biology processes, and their design is an exciting area of ongoing engineering endeavor. To understand biology and unlock future technology, we must understand the principles behind molecular machine operation.

Molecular machines are notably distinct from the macroscopic machines that have been honed by engineers for centuries. 
The high-level differences result from their nanometer scale and material composition.
These, and a few other \emph{stylized facts}~\footnote{Borrowing from economics, a \emph{stylized fact} is a broad generalization that summarizes empirical data but may not capture all cases~\cite{StylizedFacts}.} about the basic features of molecular-machine operation, provide sufficient constraints to point to relevant simple models and governing physical limits. 
Stochastic thermodynamics~\cite{Seifert:2012es,Jarzynski:2011:AnnRevCondMattPhys} is a fruitful and increasingly promising framework within which to understand their operation.

Free energy transduction is a key element in describing how molecular machines work. Because useful molecular machines are in contact with nonequilibrium reservoirs of free energy and material, they are not in equilibrium; the second law of thermodynamics provides guidance, indicating that free energy must be consumed to do useful things.

Here, we review theoretical and computational explorations of nonequilibrium free energy transduction in molecular machines.
Initial sections of this review lay the groundwork for understanding free energy transduction in molecular machines. 
\S\ref{sec:types} introduces molecular machine classes and modes of free energy transduction, concluding in \S\ref{sec:caseStudies} with the introduction of two model systems that will be frequent illustrating examples throughout this review.
\S\ref{sec:stylizedFacts} summarizes stylized facts which point to considerations when modeling molecular machine behavior.
\S\ref{sec:models} outlines the basic elements of popular modeling approaches and \S\ref{sec:physicsBackground} the kinetic and thermodynamic concepts that figure prominently in such models. 
\S\ref{sec:performance} describes common measures of performance that quantitatively describe the functionality of molecular machine operation, 
\stkout{as well as} 
example values from various molecular machines,
{\fix as well as how machine characteristics can lead to improved performance.}
\stkout{Subsequent sections} 
{\fix Sub-sections of \S\ref{sec:performance}} discuss in more detail several frameworks for analyzing and computing molecular machine free energy transduction, {\fix with focuses on} 
\stkout{while \S 
 examines} 
how free energy is consumed to generate autonomous and reliable directed machine behavior,
\stkout{\S 
 explores} 
molecular machine features that lead to fast and efficient operation, 
\stkout{as well as}
{\fix and}
trade-offs between various performance measures.
\S\ref{sec:deterministicDriving} covers the energetic costs incurred by control protocols. 
Although we focus primarily on naturally evolved biomolecular machines, \S\ref{sec:synthetic} briefly discusses insights gained from and possible applications to synthetic molecular machines.  
\S\ref{sec:emerging} outlines several areas of emerging and future interest. 
\S\ref{sec:conclusion} finishes with some concluding thoughts.

We focus on recent literature, though of necessity we refer back to earlier work to introduce relevant frameworks, models, and questions.
We also refer the reader to an excellent set of reviews in previous years on various aspects of molecular machine operation and modeling from a theoretical perspective~\cite{Julicher:1997dq,qian99,Bustamante:2001us,Hong:2005ev,Kolomeisky:2007wh,Astumian:2011eq,Seifert:2012es,Kolomeisky:2013ir,Chowdhury:2013bfa,Chowdhury:2013fn,Qian:2016ef,Hoffmann:2016gb,Goychuk:2016hv,Pezzato:2017bz,Brown:PiC,Astumian:2018fa}.

\section{Types of molecular machines}
\label{sec:types}
Biomolecular machines fulfill a wide variety of cellular roles~\cite{Chowdhury:2013fn}. 
At their most fundamental, each type of molecular machine transduces one nonequilibrium store of free energy into another, converting among mechanical energy, electrical energy, chemical energy, and low-entropy distributions both across space and among chemical species (small molecules and biopolymer sequences).

Transport motors~\footnote{In this review, we use \emph{machine} to describe any free energy transducer but reserve \emph{motor} for machines with a functional translational or rotational motion.} (such as kinesin{\fix~\cite{Hirokawa:2009wo}}, dynein~\cite{reck-peterson18}, and myosin{\fix~\cite{Sellers:2000cw}}) haul cellular cargoes (such as organelles or chromosomes) along cellular filaments~\cite{Howard}, thereby transducing chemical potential differences (often between ATP and ADP+$\text{P}_{\text{i}}$) into directed mechanical forces and ultimately into spatial concentration differences.  
Translocases (such as the $\phi$29 packaging motor~\cite{rao08} or \stkout{DNA helicase} {\fix nucleic-acid helicases~\cite{Singleton:2007p27207}}) pull biopolymers (during packaging or unwinding), thereby transducing chemical potential differences into mechanical forces and ultimately high pressures (packaging motors) or redistribution across free energy barriers.
Muscle motors (such as myosin in actomyosin fibers{\fix~\cite{alberts14}}) provide motion in muscles, thereby transducing chemical potential differences into linear movement against loads (i.e.\ performing work).
Pumps (such as $\ce{Na^+}/\ce{K^+}$-ATPase~\cite{jorgensen03} and electron-transport complexes I, III, and IV in cellular respiration~\cite{alberts14}) push small molecules across membranes, thereby transducing chemical potential differences between reactants and products into concentration differences of another chemical species across membranes. 
Polymerases (DNA polymerase{\fix~\cite{Berdis:2009fb}}, RNA polymerase{\fix~\cite{PARKER20011746}}, and the ribosome{\fix~\cite{Ramakrishnan:2002ul}}) add monomers to the end of biopolymers, thereby transducing chemical potential differences into low-entropy distributions of polymer sequences. 
Rotary motors (such as $\Fo\F1$-ATP synthase{\fix~\cite{Junge:2015fo}} and the bacterial flagellum{\fix~\cite{Berg:2008ia}}) transduce electrochemical differences across membranes into rotation against a torque (and thus perform work). 

Molecular machines often work as part of tightly or loosely coupled larger assemblies to accomplish their functions: transport motors can work together (and antagonistically, when oppositely directed) to transport cargoes{\fix~\cite{McLaughlin:2016jc,Hancock:2014fc}}; $\Fo\F1$-ATP synthase is an intimate assembly of two oppositely directed rotary motors{\fix~\cite{Junge:2015fo}}; polymerase holoenzymes include core polymerases, helicases, and error-checking apparatus~\cite{Baker:1998tx,alberts14}; and actomyosin fibers consist of precise spatial arrangements of myosin motors~\cite{alberts14}.


\subsection{Case studies}
\label{sec:caseStudies}
When illustrating the concepts in this review, we focus on two molecular machines, ATP synthase and kinesin. 
Despite this special emphasis, the ideas in this review apply beyond these two examples to other machines, and to systems sometimes not considered as molecular machines~\cite{Slochower:2018kx}, such as catalytic enzymes~\cite{Bustamante:2004fo}. 

\begin{figure}[ht]
\centering
\begin{tabular}{cc}
\includegraphics[width=1.8in]{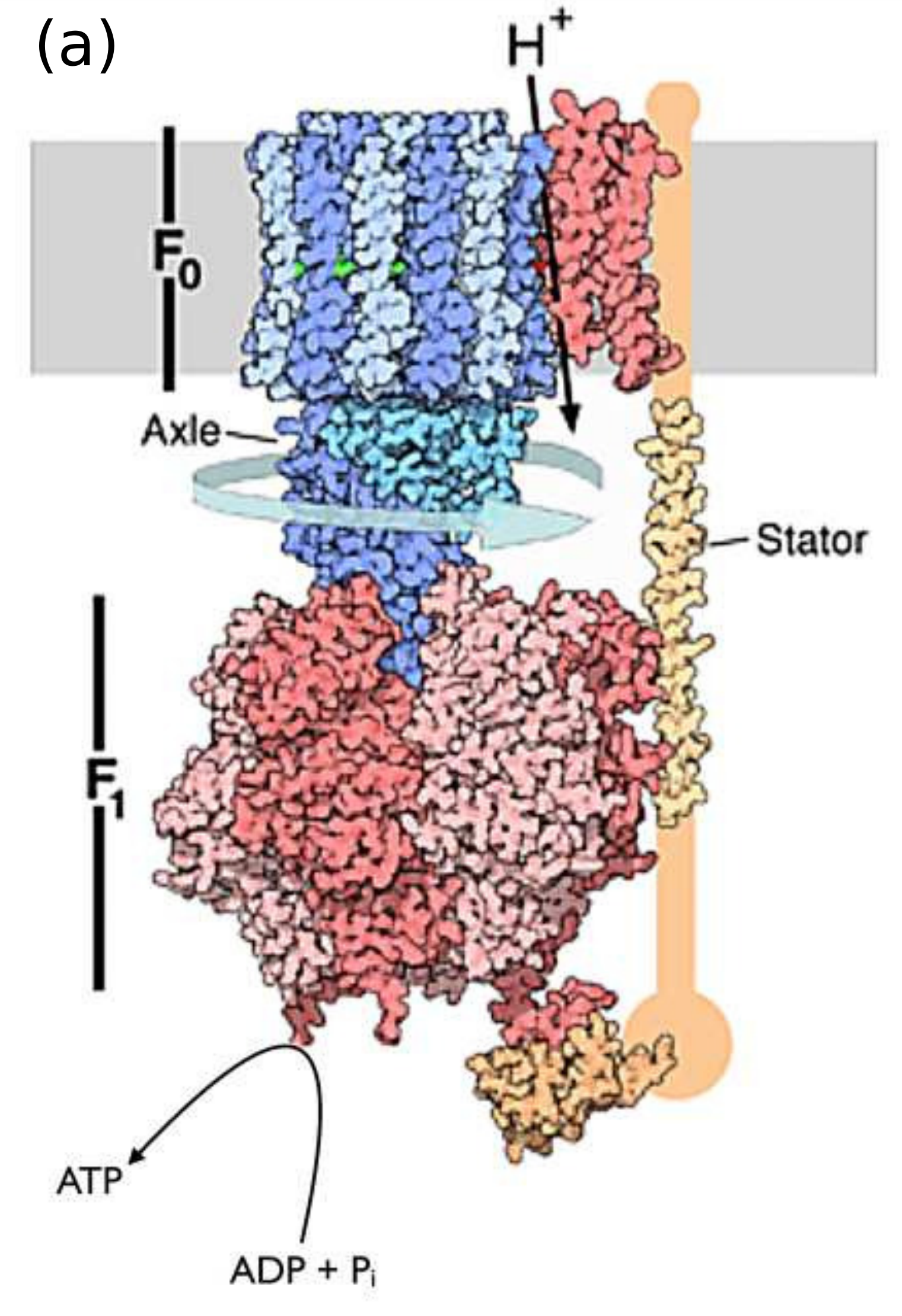} &
\includegraphics[height=2.7in]{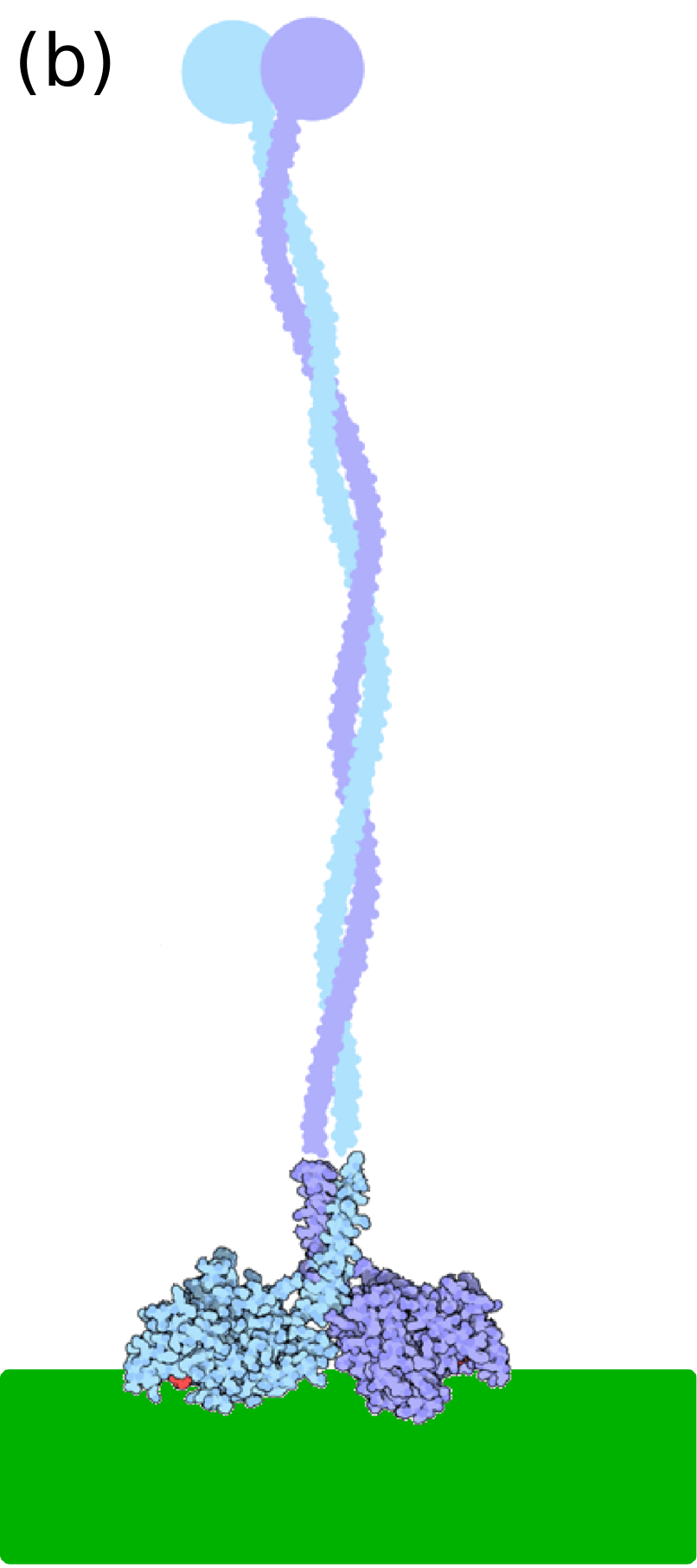}
\end{tabular}
\caption{
\label{fig:atpsynthase}
{\bf Model systems.}
(a) ATP synthase {\fix ($\sim$10 nm in diameter~\cite{Boyer:1997wi})}, a composite rotary motor. 
The membrane-embedded $\Fo$ motor (top, blue) couples proton flow across the membrane (down the electrochemical gradient) to rotation of the central axle or crankshaft. 
The cytoplasmic $\F1$ motor (bottom, red) couples this rotation to synthesis of ATP from ADP and phosphate. 
(b) Representation of the crystal structure of kinesin, a motor that walks along microtubules (shown in green), towing cellular cargo. 
\stkout{The bottom is} 
The {\fix $\sim$7nm-diameter~\cite{Kull:1996jg}} motor domain {\fix at the bottom} (where microtubule binding and ATP hydrolysis occur) 
\stkout{and the top is}
{\fix is connected via a $\sim$80-nm linker~\cite{Jeney:2004hc} to}
the cargo-binding domain {\fix on top} (shown schematically). 
Both images adapted from the Protein Data Bank~\cite{pdb00,rastogi99,gibbons00,wilkens05,delrizzo02,kozielski97}.}
\end{figure}

ATP synthase couples transport of hydrogen ions down their gradient to synthesis of ATP from ADP and phosphate, against a chemical-potential difference favouring ATP hydrolysis{\fix~\cite{Junge:2015fo}}.
Though ATP synthase is a large and intricate molecular complex, communication is mediated through a relatively simple mechanical coordinate, the rotational angle of a crankshaft connecting the integral membrane $\Fo$ subunit to the soluble $\F1$ subunit (Fig.~\ref{fig:atpsynthase}).
Single-molecule studies of ATP synthase typically {\fix remove} \stkout{excise} the $\Fo$ subunit, attach an experimental handle (e.g., a single magnetic bead {\fix or dimeric beads}) to the crankshaft {\fix attached to the $\F1$ subunit}, and monitor or force rotation using a magnetic trap{\fix~\cite{Rondelez:2005be} or electrorotation}~\cite{toyabe11}.
Such experiments suggest that $\F1$ can approach near 100\% efficiency~\cite{Yasuda:1998:Cell,kinosita00,Rondelez:2005be,toyabe11}.

Kinesin-1 (hereafter kinesin) is a transport motor that walks toward the plus end of microtubules, powered by ATP hydrolysis~\cite{Schnitzer:1997bm,Coy:1999ug}. 
Kinesin takes discrete 8 nm steps~\cite{svoboda93} as its two `heads' alternate binding in a hand-over-hand fashion along the microtubule~\cite{yildiz03}. 
These heads contribute much of the activity necessary for kinesin walking, including binding the microtubule, hydrolyzing ATP, and gating their enzymatic activity to provide directed motion~\cite{andreasson15}. 
The two heads are connected to a `neck linker', which provides important conformational changes as part of the kinesin stepping cycle~\cite{tomishige06,zhang17}. 
The neck linker is also connected to a longer coiled coil which ends in a cargo-binding domain~\cite{cockburn18} (Fig.~\ref{fig:atpsynthase}). 
In addition to its main forward stepping pathway, kinesin can take backward steps (requiring ATP hydrolysis) and engage in futile cycles that hydrolyze ATP without taking a step in either direction~\cite{clancy11}. 
Kinesin can take many forward steps before detaching from a microtubule~\cite{toprak09}. 
By attaching a bead to kinesin, an optical trap can manipulate kinesin by applying force as it walks~\cite{clancy11}.

\section{Stylized facts}
\label{sec:stylizedFacts}
Here we outline stylized facts about the unfamiliar, often counter-intuitive, physical setting for molecular machines.
Overall, the considerations in this section suggest that nonequilibrium and statistical approaches, applied to isothermal, overdamped, thermodynamically consistent models that properly account for mechanical and chemical degrees of freedom, are central to understanding the behavior and design of molecular machines.

Molecular machines are of necessity isothermal machines. Any temperature gradient sufficiently large and on sufficiently short length scales to power meaningful motion of a molecular machine would relax long before the machine could complete a cycle~\cite{Baffou:2014kq,Hoffmann:2016gb}. 

These nanometer-sized objects, composed of relatively soft protein material, have energy scales comparable to the thermal energy $\kT$ at ambient temperature, so significant stochastic fluctuations are omnipresent~\cite{Astumian:2007cd,yanagida08}. Similar to pollen grains diffusing in water~\cite{einstein05}, the components of molecular machines (unlike macroscopic machines) experience large stochastic fluctuations as they are constantly jostled by collisions with the surrounding medium (typically water or other proteins). 
Thus, even a driven molecular machine will move in a given direction only on average, with pauses, back steps, off-pathway (side) steps, and so on~\cite{isojima16}. 
These stochastic fluctuations are another manifestation of the noise that frictionally damps molecular machine motion: at equilibrium, the frictional damping and stochastic kicks are tightly connected by the fluctuation-dissipation theorem~\cite{Chandler1987a}.~\footnote{Being out of equilibrium breaks the equilibrium fluctuation-dissipation theorem, but even out of equilibrium, the two behaviors are not unrelated: various nonequilibrium fluctuation-dissipation theorems have been derived, especially in nonequilibrium steady states{\fix~\cite{Harada:2005db}}~\cite{Seifert:2010ba,Verley:2011kb}.}

The typical length and velocity scales of molecular machines place them in the regime of low Reynolds number~\cite{purcell77}, $Re \equiv vL/(\mu/\rho)$, for velocity $v$, characteristic linear dimension $L$, viscosity $\mu$, and mass density $\rho$. Kinesin has maximum speed $\sim1 \, \mu\text{m}/\text{s}$ and size $\sim10 \, \text{nm}$, and room-temperature water has viscosity $\mu \sim 10^{-3} \, \text{Pa}\cdot\text{s}$ and an approximate density $10^3 \, \text{kg}/\text{m}^3$, suggesting a Reynolds number of $\text{Re} \sim 10^{-8}$~\cite{Holzwarth:2002jq}. Thus viscous (frictional) forces dominate inertial forces, so the motion of molecular machines is completely overdamped: Any instantaneous direction of motion is rapidly randomized by energetic collisions with the machine's surroundings. As a result, a molecular machine rapidly `forgets' its direction of motion. Unlike a macroscopic machine, a molecular machine cannot rely on inertia to carry it through any particular stage of its cycle. The average motion of such nanoscale objects persists only as long as something continues to `push'. Essentially, any machine `velocity' emerges only from an imbalance of forward vs.\ backward steps, not from any physically meaningful instantaneous velocity.

Molecular machines operate in the crowded environment of the cellular interior~\cite{ellis01,zhou08,hofling13}. Water has a viscosity of $10^{-3}\text{~Pa}\cdot\text{s}$, while viscosity measurements inside the cell reach as high as $10^3\text{ Pa}\cdot\text{s}$, a million-fold higher~\cite{caragine18}. This higher intracellular viscosity is due to \emph{macromolecular crowding}, i.e.\ the high concentrations of macromolecules which occupy 10-40$\%$ of the cell volume~\cite{theillet14}. This suggests that the Reynolds number experienced by biomolecular machines is substantially smaller than even the low value they would experience if in water. 

Molecular machines typically operate cyclically,~\footnote{There are exceptions, `one-shot' machines such as the spasmoneme~\cite{Mahadevan:2000ji}.} allowing repetition of a task, with stochastic behavior on shorter timescales averaging to more reliable output over longer timescales. These machine cycles involve events with both chemical and conformational changes. Chemical reactions and ligand binding/unbinding events proceed essentially instantaneously compared to conformational relaxation time scales, but waiting times for chemical reactions can exceed conformational relaxation time scales.

Molecular machines are often assembled from many components. These components are strongly coupled, and interact with different aspects of their environment. The emerging field of stochastic thermodynamics of strongly coupled systems provides promising frameworks~\cite{Strasberg:2016dr,Strasberg:2017it,Seifert:2016ik,Jarzynski:2017hf} to investigate free energy transduction among the various components of machines. 

The elasticity of the coupling between different machine components allows energy to be elastically stored and perhaps accumulated at an interface before release, essentially smoothing any mismatches between interacting components. 
This permits the kinetic decoupling of the behavior of two interacting machines~\cite{Junge:2015fo}. For example, in $\Fo\F1$-ATP synthase, elastic energy storage may permit the transmission of energy between components that have different periodicities, such as $\Fo$ and $\F1$~\cite{Junge:2001wu, Sielaff:2008cn, Junge:2009gr, Okuno:2010eva, Wachter:2011cp, Czub:2011ci, Okazaki:2015dr}.


\subsection{{\fix Microscopic reversibility and nonequilibrium driving}}
Molecular machines, with overdamped motion and jostled by fluctuations, can nonetheless complete the stages composing their operational cycles. Microscopic reversibility~\cite{onsager31,astumian12} dictates that for every trajectory that completes a molecular machine stage in the forward (functional) direction, there must be a corresponding physically realizable reverse (dysfunctional) trajectory. Any forward step must be matched by the potential for a backward step, however unlikely. Thus the operation of any molecular machine is in principle microscopically reversible, capable of reversing any free energy conversion to run in the opposite direction, though the reverse operation may be so unlikely as to be unobserved in a given experiment. \stkout{Thus} {\fix Hence} 
is of necessity a probabilistic phenomenon, not deterministic. This (mechanical and chemical) reversibility has actually been experimentally demonstrated~\cite{Kolomeisky:2013ir} for $\F1$-ATPase{\fix \cite{Rondelez:2005be}}~\cite{toyabe11}, the full $\Fo\F1$-ATP synthase{\fix \cite{Diez:2004bv}}, and an isolated stage of the kinesin cycle~\cite{hackney05}~\footnote{Note that some other findings of `reversibility' (e.g. in myosin V~\cite{Sellers:2010ch,Gebhardt:2006ho} and kinesin~\cite{Carter:2005bf}) demonstrated mechanical reversibility without establishing reversal of the chemical reactions.}.

At thermal equilibrium, microscopic reversibility and the resulting detailed balance~\cite{vanKampen} requires zero net flux between different states. Hence molecular machines at equilibrium are not functional and do no useful work: a transport motor at equilibrium is as likely to take backward steps as forward steps; $\Fo\F1$-ATP synthase at equilibrium is as likely to hydrolyze as synthesize ATP.

This immediately implies that molecular machines must be out of equilibrium in order to be functional. It is fundamental to the operation of molecular machines---and the directed behavior they must achieve---that they operate out of equilibrium. The ability of a molecular machine to achieve directed motion relies on driving by nonequilibrium forces, which necessarily dissipate free energy. 

\stkout{
Molecular machine operation with average net progress must be out of equilibrium.
Accordingly, the directed behavior of molecular machines requires consumption of free energy.}

\begin{figure}[ht]
\centering
\includegraphics[width=3in]{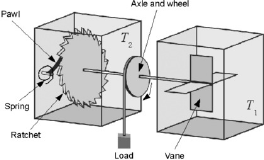}
\caption{
\label{fig:ratchetpawl}
{\bf Feynman's ratchet and pawl.} 
A vane rotationally fluctuates from impact with diffusing gas molecules. The vane rotates a ratchet (with asymmetric teeth) which has a pawl to prevent rotation in the `wrong' direction when held down by a spring. 
If the vane and ratchet rotate in the correct direction, 
the device can do useful work by lifting a load.
Reproduced with permission from Ref.~~\cite{tu08}. Copyright 2008 Institute of Physics.}
\end{figure}

This requirement of free energy for directed behavior is illustrated by Feynman's ratchet and pawl~\cite{feynman66} (Fig.~\ref{fig:ratchetpawl}), which describes a wheel with asymmetric teeth that has its rotation in one direction (counterclockwise, without loss of generality) limited by a pawl. The wheel is coupled to a vane to introduce fluctuations, with the stochastic impact of gas molecules occasionally driving the wheel clockwise. This scenario appears to violate the second law of thermodynamics, as thermal gas molecule fluctuations drive directed rotation of the wheel. The second law in fact holds because the same fluctuations that drive rotation in the clockwise direction also disengage the pawl, allowing counterclockwise motion, and thus as a whole the thermal fluctuations do not produce net average rotation in either direction~\cite{jarzynski99}.

There are two requirements to rectify thermal fluctuations into directed motion and hence, potentially into mechanical work~\cite{Astumian:2011eq}: spatial asymmetry and nonequilibrium driving~\cite{Hoffmann:2016gb}. 
Nonequilibrium driving provides temporal asymmetry to molecular machine dynamics, 
{\fix and spatial asymmetry permits directed response to free energy input} 
\stkout{but without spatial asymmetry, the molecular machine cannot respond to free energy input in an asymmetric (directed) fashion}. 

The free energy consumed by molecular machine operation, which is not transduced into another store of free energy, is known as dissipation. 
Although some molecular machines are quite efficient, and convert between forms of free energy such that most of the free energy from fuel is later available for some other process, other machines dissipate much of the free energy from fuel. For example, in \stkout{some scenarios kinesin can dissipate up to 80\% of} {\fix physiologically relevant conditions, kinesin dissipates most of its} input free energy~\cite{Ariga:2018jj}.
This dissipated 
free energy provides forward bias (see \S\ref{sec:freeEnergy}).
At macroscopic scales, this free energy cost of directionality is also present, but can be much smaller in comparison to the total free energy input{\fix~\cite{machta15}}. 

In contrast to heat engines driven by temperature differences, biomolecular machines are isothermal and instead driven by nonequilibrium chemistry, typically from nonequilibrium concentrations of reactants and products or gradients maintained across membranes.
Many machines are driven by ATP hydrolysis; in physiological contexts, concentrations of ATP and its hydrolysis products ADP and $\rm P_{\rm i}$ are maintained out of equilibrium so that they provide a 20-$\kT$ driving force~\cite{CellBioByNum}, dwarfing the scale of thermal fluctuations.

The directionality imposed by out-of-equilibrium concentrations of chemical species depends on the likelihood of particular species binding to the machine. Only those kinetic steps involving binding of chemical species are sensitive to the concentrations of those species. In particular, conformational changes that don't change the binding state of a machine are not biased by chemical-potential driving forces~\cite{astumian12}.

\section{Models}
\label{sec:models}

Given the qualitative regimes describing molecular machines in \S\ref{sec:stylizedFacts}, many researchers have developed simple models for molecular-machine energetics and dynamics to gain \stkout{physical} understanding.
To be analytically tractable, models of molecular machines typically have few degrees of freedom. These degrees of freedom include mechanical or conformational variables, binding and unbinding reactions (to ligands or cytoskeletal filaments), and chemical degrees of freedom.

\subsection{Continuum models}
Molecular machine dynamics can be modeled as diffusion on an energy landscape with a continuous state space. For a given generalized machine coordinate $\vec{x}$, the energy is $V(\vec{x})$. The machine moves on the energy landscape with a rapidity depending on the diffusivity $D(\vec{x})$, which is often assumed to be uniform.

The system dynamics are typically analytically or numerically solved using one of two primary approaches: solving for the temporal evolution of the probability distribution using the Fokker-Planck equation~\cite{zhang09}, or simulating realizations of individual trajectories using Monte Carlo or Langevin methods~\cite{vanKampen,handStochasticMethods,Risken1996,frenkelSmit,Sivak:2013:PhysRevX,Sivak:2014:JPhysChemB}.

\begin{figure}[ht]
\centering
\includegraphics[width=4.5in]{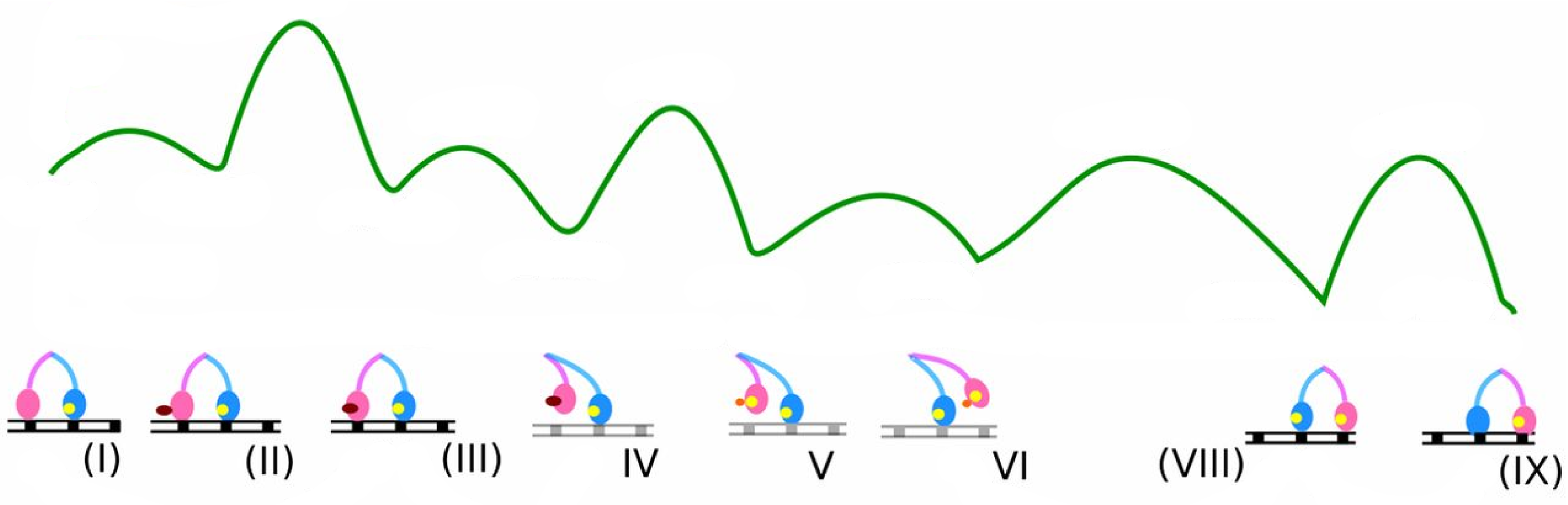}
\caption{
\label{fig:continuouslandscape}
{\bf Continuous free energy landscape.} 
Top: free energy landscape for a single step of myosin-V on an actin filament. Bottom: Schematic progression of myosin-V states (sequentially labelled with Roman numerals) aligned with the location of corresponding metastable states along the free energy landscape. 
Adapted from Ref.~\cite{mukherjee17}.}
\end{figure}

Often the multidimensional state space of the molecular machine is reduced to a single coordinate $x$, along which reaction progress in the `forward' direction is considered (Fig.~\ref{fig:continuouslandscape}). Other degrees of freedom (presumed to relax faster) are averaged over using a timescale-separation argument~\cite{vanKampen:1985p64981} to arrive at the free energy $V_{\rm eff}(x)$ and diffusivity $D_{\rm eff}(x)$ along this single coordinate.

Multiple model types can describe the generation of directed behavior in molecular-machine models. Here we outline two such models.



\begin{figure}[ht]
\centering
\includegraphics[width=3in]{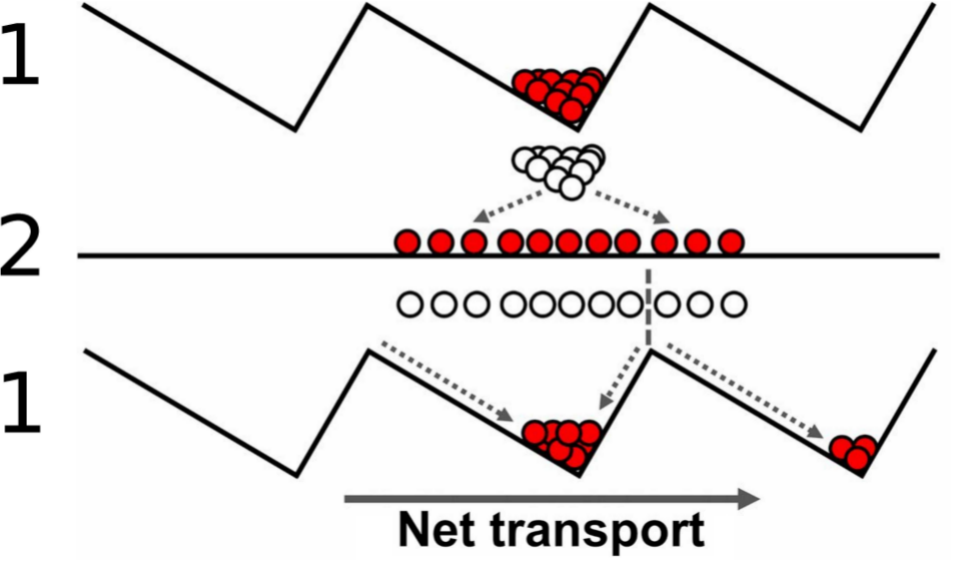}
\caption{
\label{fig:flashingRatchet}
{\bf A flashing ratchet} generating directed behavior through alternation between two potentials. In the sawtooth potential (1), the machine generally localizes to the bottom of the free energy wells (red particles represent ensemble of final machine positions in potential). When the potential switches to the flat potential (2), the white particles represent the ensemble of initial machine positions, which then diffuse isotropically (final machine positions again represented by red particles). When the potential flashes back to the sawtooth potential (initial machine positions shown as white particles), a machine that has reached a neighboring free energy well tends to relax to the bottom of that well, rather than returning to the initial well. 
Diffusion is isotropic, but because the free energy wells are asymmetric, the machine is more likely to reach the basin of a neighboring well to the right rather than to the left: On average, the machine makes directed progress.
Adapted from Ref.~\cite{kedem17}.}
\end{figure}

One is a \emph{flashing potential}, that instantaneously switches between multiple free energy landscapes~\cite{Doering:1995iy} (Fig.~\ref{fig:flashingRatchet}). The alternating potentials are chosen such that, on average, this switching will drive the machine in one direction. A typical scenario involves two features: energy landscapes whose periodic features are not aligned,
and sufficiently long wait times between switching to allow the machine time to find an energy minimum. Figure~\ref{fig:flashingRatchet} shows how isotropic diffusive spreading can be rectified by asymmetric energy wells as part of a flashing potential. 

Another standard model is to periodically repeat the entire landscape representing the machine cycle, with an overall free energy drop over the course of the entire cycle, such that the landscape is `tilted'~\cite{mel1991kramers}. As the machine diffuses on the free energy landscape, it will tend towards lower free energy, generating directed behavior in that direction~\cite{Risken1996}.

\subsection{Discrete-state models}
\begin{figure}[ht]
\centering
\includegraphics[width=3.5in]{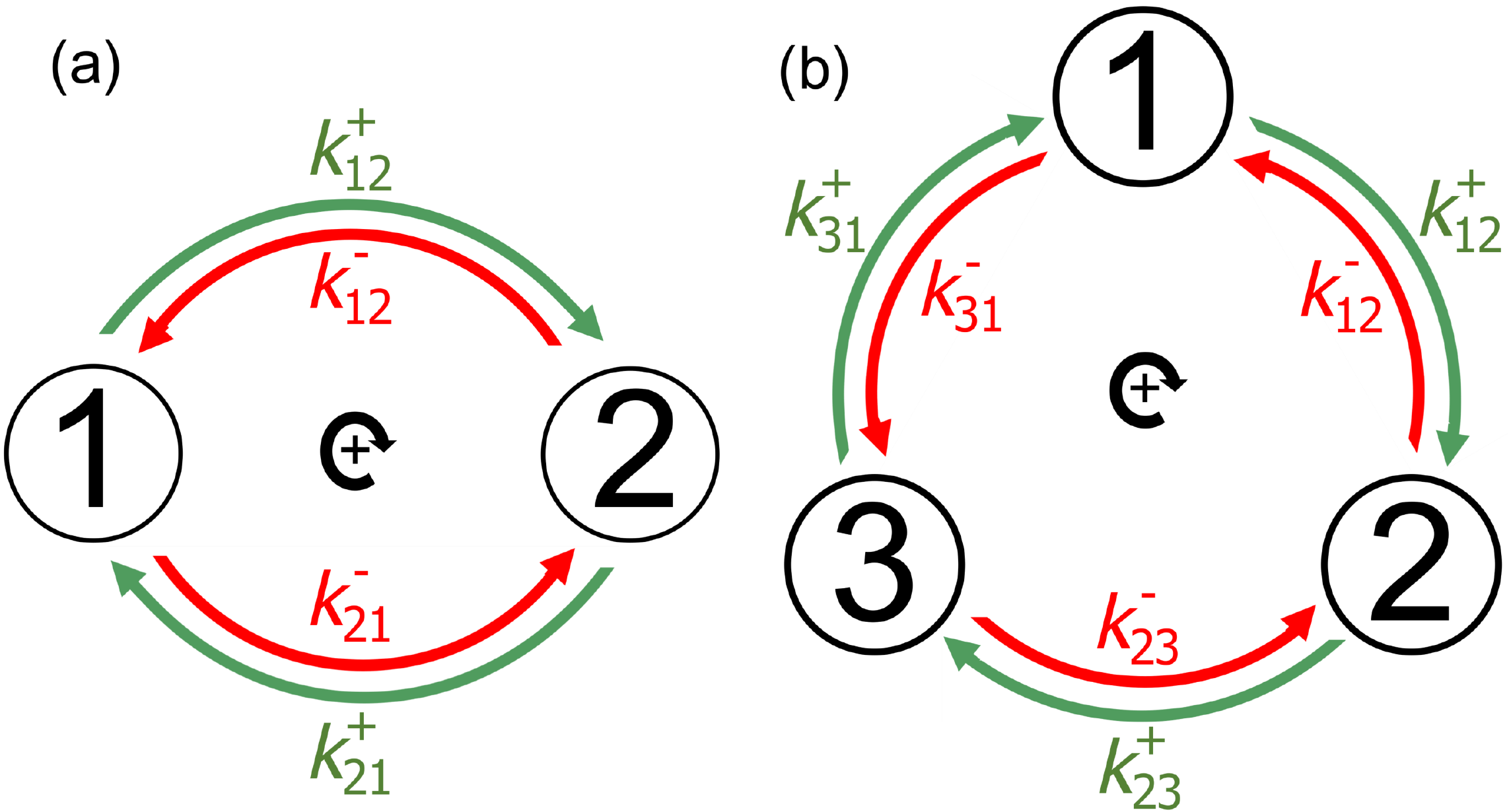}
\caption{
\label{fig:discretecycles}
{\bf Discrete-state models} with (a) two and (b) three states per cycle. Transitions occur in both the forward (green arrows) and reverse (red arrows) directions for each pathway, with respective rate constants $k_{ij}^+$ and $k_{ij}^-$. 
Reproduced from Ref.~\cite{Brown:2018ip}. Copyright 2018 American Chemical Society.}
\end{figure}

Alternatively, the state space of molecular machines is often divided into a set of discrete states~\cite{qian97} (Fig.~\ref{fig:discretecycles}). This discrete-state description is motivated by the commonly encountered situation of large barriers separating compact metastable macrostates, when a timescale separation exists between relatively long timescales spent within a given macrostate and relatively short timescales to actually make the transition between macrostates~\cite{Challis:2018kp}. 

$k_{ij}^+$ is the transition rate constant in the forward direction from state $i$ to state $j$; $k_{ij}^-$ is the reverse transition rate constant from state $j$ to state $i$ along the same transition path. This description allows for physically distinct pathways between two states that, though they begin in the same state and finish in the same state, have distinct effects on the environment\stkout{, e.g.\ producing or consuming a small molecule}. {\fix For example, in Fig.~\ref{fig:discretecycles}, $k_{12}^+$ and $k_{12}^-$ could be the respective rates of phosphorylation and dephosphorylation by ATP-driven enzymatic catalysis, and $k_{21}^+$ and $k_{21}^-$ the respective rates of undriven non-enzymatic dephosphorylation and phosphorylation.}

A continuous model can incorporate more details in the free energy landscape and allow the assessment of behavior during transitions, but discrete-state models can more computationally efficiently generate trajectories and dynamic probability distributions.

\subsection{Hybrid models}
\label{sec:hybrid}
\begin{figure}[ht]
\centering
\includegraphics[width=3.5in]{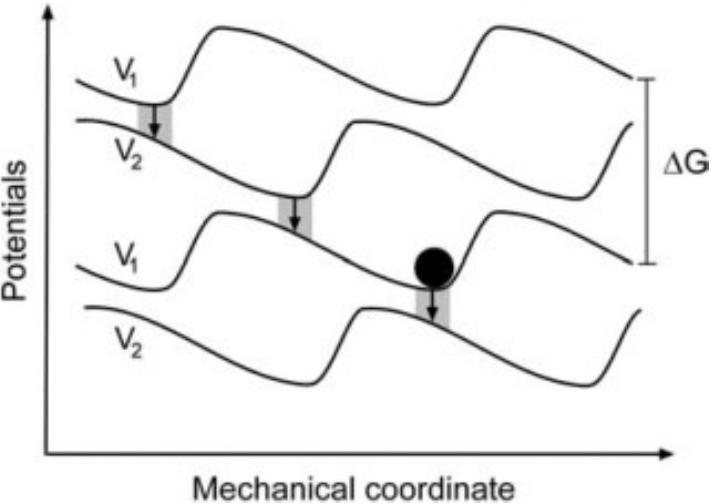}
\caption{
\label{fig:hybrid}
{\bf Hybrid model.} Schematic continuous free energy landscapes connected by discrete transitions between distinct landscapes from a particular location on each landscape. 
Reproduced with permission from Ref.~\cite{xing05:ContinuumDiscrete}. Copyright 2005 The Biophysical Society.}
\end{figure}

Continuous- and discrete-state molecular machine models can be combined in models that feature discrete transitions between multiple continuous free energy landscapes~\cite{Xing:2005ft} (e.g.~Fig.~\ref{fig:hybrid}). This combination permits the use of discrete or continuous descriptions of different degrees of freedom, as appropriate. For example, though the waiting time between chemical reactions can be long, the actual transition time is typically much faster than characteristic time scales for protein conformational rearrangements, so a chemical reaction can be modeled as a discrete transition between continuous mechanical landscapes.

A flashing potential is an example of a hybrid model that typically transitions between potentials independently of the machine state.
Instead, the potentials alternate on a set schedule.

\subsection{Atomistic or coarse-grained molecular models}
More detailed computational models either explicitly represent all atoms~\cite{Ovchinnikov:2011bb,Okazaki:2015dr,Czub:2017cq}{\fix~\cite{Dai:2017jm}} or coarse-grain individual atoms into larger effective particles~\cite{Pu:2008hn,Mukherjee:2015gn}{\fix~\cite{Isaka:2017gh}}. The number of degrees of freedom and complexity of force fields preclude any analytic treatments, so these models are solved using discrete integration of equations of motion, either deterministic or stochastic~\cite{frenkelSmit}. Such models can incorporate significantly more molecular detail and potentially resolve the ramifications of such details on machine performance, but their increased computational cost limits the accessible simulation time scales and reduces the breadth of statistical sampling.

\section{Background quantitative concepts}
\label{sec:physicsBackground}
Given the roles of molecular machines (\S\ref{sec:types}), the qualitative environment they encounter (\S\ref{sec:stylizedFacts}), and the models used to describe their dynamics (\S\ref{sec:models}), we now outline some \stkout{physical} quantities and properties important to understanding molecular machine behavior.

\subsection{Flux}
\label{sec:flux}
The macrostate of a molecular machine is often described using the probability distribution across microstates. For a discrete-state model, the probability of occupying each discrete microstate $i$ is $P_i$, and the directed (one-way) probability flux from state $i$ to state $j$ along a given pathway is
\begin{equation}
J_{ij}^+ = k_{ij}^+P_i \ ,
\end{equation}
and from state $j$ to state $i$ is
\begin{equation}
J_{ij}^- = k_{ij}^-P_j \ .
\end{equation}
The net flux, the net flow of probability from state $i$ to state $j$ along a given pathway, is the difference of the directed fluxes,
\begin{equation}
J_{ij} = J_{ij}^+ - J_{ij}^- = k_{ij}^+P_i - k_{ij}^-P_j \ .
\end{equation}

A system at equilibrium satisfies \emph{detailed balance},
\begin{equation}
k_{ij}^+ P_i = k_{ij}^- P_j \ ,
\end{equation}
where for each pathway $i\leftrightarrow j$ the `forward' flux $J_{ij}^+$ from $i$ to $j$ is exactly balanced by the `reverse' flux $J_{ij}^-$ from $j$ to $i$,
\begin{equation}
\label{eq:db1}
J_{ij} = 0 \ .
\end{equation}

\subsection{Nonequilibrium steady state}
For constant external conditions, a machine can reach a \emph{nonequilibrium steady state} (NESS), when for each state the total incoming and outgoing fluxes balance,
\begin{equation}
\label{eq:ness}
\sum_j (J_{ij}^+ - J_{ij}^-) = 0 \ , \ \forall i\ .
\end{equation}
The distinction of a NESS from equilibrium is that the state probabilities do not necessarily satisfy the detailed balance condition~\eqref{eq:db1}, for which each and every individual flux balances. Although in a NESS the state probabilities remain constant over time, these probabilities do not generally satisfy the Boltzmann distribution~\cite{trepagnier04}. We will primarily consider motors in a NESS, as any long-term averages should be dominated by NESS behavior and only marginally affected by transient behavior while relaxing to a NESS.

For a discrete-state model which has reached a NESS, the flux can be calculated using Hill's diagrammatic method~\cite{hill77}. This approach gives the NESS flux for a two-state cycle (Fig.~\ref{fig:discretecycles}a),
\begin{equation}
\label{eq:ness2states}
J_{\text{2-state NESS}} = \frac{k_{12}^+k_{21}^+ - k_{12}^-k_{21}^-}{k_{12}^+ + k_{12}^- + k_{21}^+ + k_{21}^-} \ ,
\end{equation}
and for a three-state cycle (Fig.~\ref{fig:discretecycles}b),
\begin{equation}
J_{\text{3-state NESS}} = \frac{k_{12}^+k_{23}^+k_{31}^+ - k_{12}^-k_{23}^-k_{31}^-}{k_{12}^+k_{23}^- + k_{12}^-k_{23}^- + k_{23}^-k_{31}^- + k_{12}^+k_{31}^+ + k_{12}^-k_{31}^+ + k_{12}^-k_{31}^- + k_{12}^+k_{23}^+ + k_{23}^+k_{31}^+ + k_{23}^+k_{31}^-} \ .
\end{equation}
Generating function methods~\cite{Koza:1999dq,Koza:2000gq,Chemla:2008fs} are a popular alternative to the diagrammatic method.  NESS fluxes for models with more states or multiple cycles can also be calculated via either method, but the expressions quickly become unwieldy.

\subsection{Free energy}
\label{sec:freeEnergy}
The dissipated free energy $\omega_{ij}$ is related to the bias of transition rate constants through the \emph{generalized detailed balance} condition:
\begin{equation}
\label{eq:bias}
\beta \omega_{ij} = \ln \frac{k_{ij}^+}{k_{ij}^-} \ .
\end{equation}
Here $\beta \equiv (k_{\text{B}}T)^{-1}$. 
For an autonomous molecular machine, $\omega_{ij}$ is the {\fix free energy dissipation} cost \stkout{(in terms of dissipated free energy)}
paid for biased forward progress at discrete transition $ij$. 
{\fix This dissipated free energy $\omega_{ij}$ is the height of the downhill drop the machine experiences over the transition, as dissipated energy is not subsequently available to perform useful work. The forward bias provided by $\omega_{ij}$ is often described as `driving' the machine.}

\subsection{Control-parameter protocol}
\label{sec:protocols}
An equilibrium ensemble can be parameterized by a control parameter $\lambda$ manipulated by an experimentalist. 
Molecular machines are often {\fix experimentally} probed by temporal variation of control parameters such as the distance between foci of optical traps~\cite{Sivaramakrishnan2013} or the rotational angle of a magnetic trap{\fix ~\cite{Rondelez:2005be}}.
A \emph{protocol} $\Lambda$ specifies a temporal driving schedule for changing the control parameter $\lambda(t)$ from initial value $\lambda_i$ to final value $\lambda_f$ in some specified time.

\subsection{Affinity}
For a molecular machine driven by constant nonequilibrium external conditions, the \emph{affinity} $\affin$ is the total amount of free energy driving forward progress per cycle, the sum of free energy dissipation along each transition~\cite{barato15,Qian:2016ef}: 
\begin{equation}
\label{eq:affinity}
\affin = \sum \omega_{ij} \ ,
\end{equation}
where $\omega_{ij}$ is given by Eq.~\eqref{eq:bias}. Molecular-machine models with multiple cycles (e.g.\ kinesin models~\cite{clancy11} with backsteps and futile cycles) have a distinct affinity for each cycle.

Affinity is equivalently related to the ratio of the product of all forward transition rate constants to the product of all reverse transition rate constants around a machine cycle~\cite{Qian:2016ef},
\begin{equation}
\label{eq:affinity2}
\beta \affin = \ln\frac{\prod_{i=1}^N k_{ij}^+}{\prod_{i=1}^N k_{ij}^-} \ .
\end{equation}

For a system experiencing only chemical driving, the maximum affinity equals (minus) the free energy change for chemical reactants converting to products. For ATP hydrolysis into ADP and inorganic phosphate, $\text{P}_{\text{i}}$, the maximum affinity is
\begin{equation}
\label{eq:ATPhydrolysisaffinity}
\affinMax = \Delta G_{\text{ATP hydrolysis}} = \Delta G_0 + \kT \ln\frac{[\text{ADP}][\text{P}_{\text{i}}]}{[\text{ATP}]} \ ,
\end{equation}
the sum of the intrinsic free energy change $\Delta G_0$ of the reaction and the logarithm of the ratio of nonequilibrium concentrations.

The maximum affinity $\affinMax$ in Eq.~\eqref{eq:ATPhydrolysisaffinity} is the free energy available for the molecular machine, both to dissipate and transduce to another free energy reservoir. The affinity in Eqs.~\eqref{eq:affinity} and \eqref{eq:affinity2} is the free energy dissipated by the molecular machine. The difference, $\affinMax-\affin$, is the free energy transduced from the fuel to another free energy reservoir, e.g.\ moving an ion from low concentration to high concentration.

\subsection{Microscopic reversibility}
\label{sec:microscopicreversibility}
In equilibrium, system dynamics must satisfy detailed balance~\eqref{eq:db1}, with the forward flux of any pathway exactly balanced by the reverse of the pathway. Implicit in the concept of detailed balance is the idea that any forward trajectory must have a corresponding reverse trajectory that is in principle possible, including when out of equilibrium~\cite{onsager31,astumian12}. This principle of \emph{microscopic reversibility} manifests itself in the construction of thermodynamically consistent models, requiring that for each forward transition there must be a corresponding reverse transition.

Although the reverse of a particular microscopic process is always possible, it is not necessarily likely. The Crooks fluctuation theorem describes the relative likelihood of forward and reverse trajectories in terms of the entropy production along the forward trajectory~\cite{crooks99},
\begin{equation}
\label{eq:crooks}
\frac{P[x(t)]}{\tilde{P}[\tilde{x}(t)]} = e^{\sigma[x(t)]/k_{\text{B}}} \ .
\end{equation}
Here $P[x(t)]$ is the probability of forward trajectory $x(t)$, $\tilde{P}[\tilde{x}(t)]$ is the probability of time-reversed trajectory $\tilde{x}(t)$, and $\sigma[x(t)]$ is the entropy produced by trajectory $x(t)$. 

According to the Crooks fluctuation theorem, the ratio of respective probabilities for a forward trajectory and the corresponding reverse trajectory exponentially increases with the entropy produced.
The Crooks fluctuation theorem~\eqref{eq:crooks} has a similar form to the generalized detailed balance condition~\eqref{eq:bias} for discrete transitions; both Eqs.~\eqref{eq:bias} and \eqref{eq:crooks} describe how increasing dissipation decreases the likelihood, relative to the forward process, of a reversed process. 

Given the centrality of fluctuation theorems (from Crooks and others) to stochastic thermodynamics, and the fundamentally stochastic nonequilibrium operation of molecular machines, many have emphasized the importance of fluctuation relations in understanding machine operation~\cite{Lacoste:2011ey}.

\subsection{Splitting factor}
\label{sec:splittingfactor}
Generalized detailed balance~\eqref{eq:bias} dictates the relation between the ratio of forward and reverse transition rates and free energy dissipation, but does not fix the absolute rates, and hence leaves unspecified how the forward and reverse rates vary with model parameters.

\begin{figure}[ht]
\centering
\includegraphics[width=3.5in]{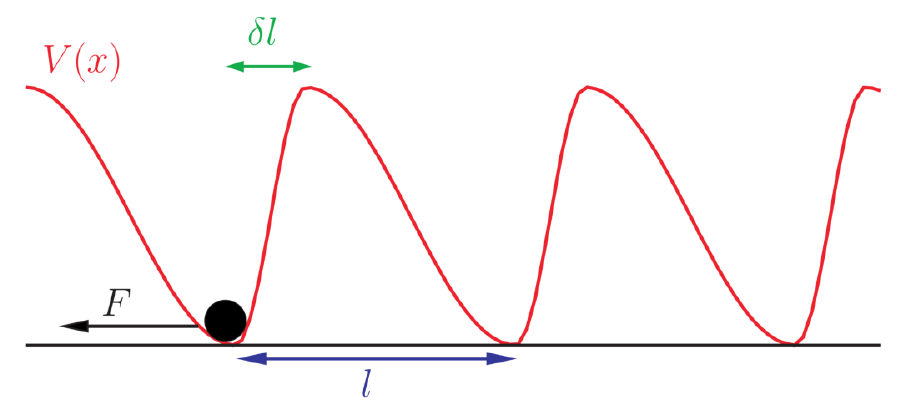}
\caption{
\label{fig:splitting}
{\bf Splitting factors.} Free energy landscape $V(x)$ for a molecular motor taking steps of size $l$ opposed by a force $F$. The transition state is positioned a fraction $\delta$ of the total step size between adjacent metastable states (i.e., a distance $\delta l$ from the state to the left and $(1-\delta)l$ from the state to the right). 
Reproduced with permission from Ref.~\cite{schmiedl08}. 
Copyright 2008 European Physical Society.} 
\end{figure}

For discrete-state models, the transition rates coarse-grain details about the underlying continuous free energy landscape. How much the dissipation affects the forward and reverse transition rates when the dissipation is varied relates to details of the free energy landscape\stkout{,}{\fix.} For example, {\fix in Fig.~\ref{fig:splitting} two states are separated by distance $\ell$, with} \stkout{the fractional distance of} the transition state {\fix a distance $\delta\ell$ in front of the rear state (to the left) and $(1-\delta)\ell$ behind the front state (to the right)} \stkout{between the two states bounding the transition}~\cite{schmiedl08,elms12,Wagoner:2016dp} \stkout{(Fig.~\ref{fig:splitting})}.

The splitting (or load-distribution) factor $\delta$ 
divides the forward bias among the forward and reverse rate constants~\cite{Brown:2018ip},
\begin{equation}
\label{eq:deltasplit}
k_{ij}^+ = k_{ij}^0 e^{\beta \delta\omega_{ij}} \ \text{and} \ k_{ij}^- = k_{ij}^0 e^{-\beta(1-\delta)\omega_{ij}} \ ,
\end{equation}
where $k_{ij}^0$ are bare rate constants that describe the inherent rate of each transition at equilibrium, combining barrier height and diffusivity along the free energy landscape. The splitting factors can also be distinct for each of multiple free energy terms for a transition, e.g.\ the `intrinsic' free energy change of the machine in a given environment and the energy change due to motion against an applied force~\cite{Brown:2018ip}. 

{\fix The applied force in Fig.~\ref{fig:splitting} acts as an additional constant-gradient 
contribution to the
energy landscape (effectively imparting a tilt to the left). If $\delta$ approaches zero, the transition state is close to the rear state, and this constant gradient adds little to the free energy hill the machine must climb to complete the forward transition (slightly decreasing $k_{ij}^+$ in Eq.~\ref{eq:deltasplit}), but substantially decreases the hill the machine must climb to reverse the transition (large 
increase
in $k_{ij}^-$ in Eq.~\ref{eq:deltasplit}). In contrast, if $\delta$ approaches 1, the transition state is close to the front state, and the constant gradient substantially increases the hill in the forward direction (substantially 
decreasing 
$k_{ij}^+$ in Eq.~\ref{eq:deltasplit}) and only slightly decreases the hill in the reverse direction (slightly 
increasing 
$k_{ij}^-$ in Eq.~\ref{eq:deltasplit}).}

\subsection{Power stroke and Brownian ratchet}
\label{sec:PSBR}
Above, we outline how splitting factors describe the effect of dissipation on forward and reverse rate constants.
A closely related concept is the mechanism by which the consumption of free energy produces directed molecular machine operation. This has frequently (including recently~\cite{Wagoner:2016dp,astumian16,Hoffmann:2016gb}) taken the form of a contrast between \emph{power strokes} and \emph{Brownian ratchets}. 
Here we survey the characteristics used to describe these two mechanisms.

A power stroke is a conformational change between two mechanical states~\cite{astumian16}, possibly driven by strain~\cite{Howard:2011kd} or stored elastic energy~\cite{Astumian:2011eq,Hoffmann:2016gb}. Mechanical transitions with power-stroke mechanisms have been described as directly driven by chemical binding, reaction, or release~\cite{Bustamante:2001us,amos08,Howard:2011kd}. These chemical processes powering the machine cycle do not need to immediately precede the power-stroke transition -- the free energy dissipation provided by chemical fuel reacting to form products can be spread over a machine cycle and used at distinct transitions. A power stroke has a transition state relatively near the pre-power-stroke state and relatively far from the post-power-stroke state~\cite{Howard:2011kd}, leading loads to primarily affect reverse rates and leave forward rates relatively unchanged~\cite{Wagoner:2016dp}. Power strokes are typically seen as decreasing free energy, but alternative voices argue that the opposite is possible~\cite{astumian16}.

Mechanisms which use free energy to rectify fluctuations that have made forward progress (by adjusting barrier heights afterwards, possibly through chemical binding or reaction), and do not perform work as part of specific conformational changes, are described as Brownian or thermal ratchets~\cite{Bustamante:2001us,amos08,Astumian:2011eq}. Strain that develops in a Brownian-ratchet mechanism is due to thermal fluctuations, rather than a specific chemical transition~\cite{Howard:2011kd}. A Brownian ratchet has a transition state relatively far from the initial state and close to the final state (in the forward direction of the machine cycle), such that loads primarily affect forward rates and leave reverse rates relatively unchanged~\cite{Wagoner:2016dp}.

Power strokes and Brownian ratchets have sometimes been presented in the literature as if they are mutually exclusive mechanisms of driving molecular machines. However, individual transitions of a molecular machine can reasonably be said to mix these two mechanisms~\cite{qian98,Hoffmann:2016gb}, and different mechanisms can operate at different stages of the same molecular machine~\cite{Bustamante:2001us}.

\stkout{At} {\fix For} the low Reynolds number {\fix dynamics} of molecular machine components, mean velocities are proportional to applied forces. Accordingly, the traditional power-stroke picture, where the development of physical strain drives a conformational change of the machine, requires sustained applied force (i.e.\ a trajectory with a continuously decreasing potential). 
\stkout{A transition} With such dynamics that are entirely driven, lacking significant diffusive contribution, {\fix a system} naturally gravitates toward a local minimum, thereby struggling to reach a specific low free energy state in a rugged landscape~\cite{Hoffmann:2016gb}. 
Although some well-studied biomolecular machine transitions may indeed be well-described by a traditional power-stroke model (e.g., muscle myosin appears to involve force on a lever arm~\cite{geeves05}), 
many conformational changes are likely to adjust the free energy landscape experienced by some other component of the molecular machine, thereby facilitating a Brownian transition.
Binding, reaction, and release of chemicals appear to often alter the free energy landscape experienced by the molecular machine, rather than directly provide forces~\cite{miyashita03,Xing:2005ft,koga06,mickler09,kondo11}. {\fix This description of machine driving, with the free energy landscape altered by 
discrete
chemical transitions, belongs to the hybrid model type 
(\S\ref{sec:hybrid}).}

\subsection{Linear response}
Linear-response theory~\cite{Chandler1987a,Zwanzig:2001} represents a popular first-order theoretical framework for near-equilibrium systems.  
It expresses system response as a linear function of nonequilibrium driving forces and dynamic control parameter changes~\cite{sivak12}.
Linear response provides a satisfactory approximation for sufficiently slow protocols or sufficiently mild driving forces.
As a first-order theory, linear response offers tractability and generality independent of many system-specific details, but sacrifices accuracy far from its applicable limits.

\section{Measures{\fix, limitations, and optimization} of performance}
\label{sec:performance}
There are many criteria to assess the performance of a molecular machine, some of which (e.g., efficiency, speed, power, and stall force) are familiar from the analysis of macroscopic machines, and some of which (e.g., precision, processivity, specificity) are distinctive to the fundamentally stochastic behavior of molecular machines. We outline a set of measures that are in principle important to a wide array of molecular machines, though the particular functional tasks and physiological context of a given machine may mean that there are no selective pressures to improve a particular measure.  

Here we also provide a sampling of empirical observations of these measures of performance, to give an overview of molecular machine capabilities and how they behave in practice. Many of these observations motivate the development of theoretical frameworks to understand \stkout{how impressive is the performance of molecular machines (i.e.,} how closely 
{\fix molecular machine performance} \stkout{do they} approach{\fix es} physical limits and how \stkout{the performance of molecular machines} {\fix that} is achieved.
There is significant interest in what machine designs lead to high performance, both because of the insight it may yield to the functioning of living systems (where the presumptive selective advantage of better performance could plausibly lead to the evolution of high-performance machines), and for the guidance it can provide to the engineering of novel molecular machines. 

Finally, we also describe how machine characteristics can lead to improved performance.

\subsection{Efficiency}
\label{sec:perfEfficiency}
{\fix Although there is not a single universal definition of} \stkout{ We define }\emph{efficiency} $\eta${\fix , it is commonly defined} as the fraction of input free energy converted to output free energy, {\fix and we first discuss this free-energetic efficiency}.

Heat engines, in contact with a hot heat bath at temperature $T_{\rm h}$ and a colder heat bath at temperature $T_{\rm c}$ (i.e.\ $T_{\rm h} > T_{\rm c}$), have efficiencies bounded by the Carnot limit, $1 - T_{\rm c}/T_{\rm h}$, which is only achieved with infinitely slow engine operation~\cite{schroeder1999thermal}. The efficiencies of molecular machines, operating isothermally and often driven by chemical potentials, are not meaningfully restricted by the Carnot limit. 
For molecular machines with tight coupling between consumption of chemical fuel and motion (i.e.\ the machine cycle and fuel consumption cannot occur separately), \stkout{an efficiency of one} 
{\fix high efficiency---even approaching unity---}can be achieved~\cite{Parmeggiani:1999twa,Bustamante:2001us,Seifert:2011cj}{\fix.}\stkout{, although flux and power will typically go to zero as no free energy dissipation remains to drive forward progress.  
Tightly coupled machines near their stall force have a high thermodynamic efficiency, approaching one, although} {\fix However,} near stall these {\fix tightly coupled} machines have low flux and power~\cite{Bustamante:2001us,Seifert:2011cj}. 
Multi-cyclic or loosely coupled machines have lower efficiency than tightly coupled machines~\cite{Seifert:2011cj}. For machines with multiple stages, the highest efficiency is achieved when the machine driving force is nearly constant~\cite{Oster:2000vd,wang02}.

Since high efficiencies achieved near stall force with very low flux do not reflect typical molecular machine operation, the efficiency is often considered at maximum power~\cite{schmiedl08}. {\fix The achievable efficiency depends on the driving regime, with low driving (low dissipation per machine cycle) distinct from greater driving.} For sufficiently low
driving 
(i.e.\ with dissipation $\ll \kT$)
the machine operates in the linear-response regime, universally achieving an efficiency at maximum power of 1/2 for tightly coupled machines~\cite{schmiedl08,VandenBroeck:2012cw}. For greater
driving, beyond the linear-response regime, efficiency can exceed 1/2{\fix~\cite{Seifert:2011cj}}, depending on the splitting factor $\delta$ (representing the transition state location)~\cite{schmiedl08,VandenBroeck:2012cw}. 
Even for a loosely coupled motor, efficiency at maximum power is maximized for a splitting factor $\delta=0$~\cite{schmiedl08}, corresponding to a transition state near the initial state rather than the final state~\cite{schmiedl08,VandenBroeck:2012cw,Wagoner:2016dp}.

\stkout{Here we discuss the efficiency~\ref{sec:perfEfficiency} with which a machine transduces one form of free energy into another.} 
Due to the stochastic dynamics of molecular machines, their efficiency is often considered as an average over many trajectories. The efficiencies of individual trajectories are subject to fluctuations, and can be negative or even exceed one, although these \stkout{macroscopically inaccessible} fluctuations {\fix(inaccessible in macroscopic systems)} are unlikely~\cite{verley14,manikandan19}.

When the only output is work against an external force $f_{\rm ext}$, the efficiency is~\cite{derenyi99,Parmeggiani:1999twa,lau07,schmiedl08}
\begin{equation}
\label{eq:efficiency1}
\eta = \frac{f_{\text{ext}}v}{J\Delta \mu} \ ,
\end{equation}
with $v$ the velocity, $J$ the flux, and $\Delta \mu$ the chemical potential change per cycle (i.e.\ the affinity $\affin$ at $f_{\text{ext}} = 0$)~\cite{Parmeggiani:1999twa,lau07}. Here, if the external force $f_{\text{ext}}=0$, the efficiency is zero.

\emph{Stokes efficiency} is {\fix a definition distinct from the free-energetic efficiency, instead} defined to include work against viscous friction \stkout{as the motor progresses}, even in the absence of an external 
conservative
force~\cite{wang02},
\begin{equation}
\label{eq:efficiency2}
\eta_{\text{Stokes}} = \frac{\zeta \langle v\rangle^2}{ J\Delta \mu + f_{\text{ext}}\langle v\rangle} \ ,
\end{equation}
with $\zeta$ the Stokes drag coefficient of the motor and cargo. $\eta_{\text{Stokes}}$ remains positive for $f_{\text{ext}} = 0$. {\fix Evaluating the Stokes efficiency allows performance comparisons between machines when the input free energy is not converted to free energy output.}

The thermodynamic uncertainty relation~\cite{barato15}~\eqref{eq:bsfano1} limits achievable molecular machine {\fix free energetic} efficiency{\fix ~\cite{pietzonka16}} \stkout{$\eta$} for {\fix a given} pulling \stkout{at} speed \stkout{$v$ against a} {\fix and} conservative force \stkout{$f_{\text{ext}}$}:
\begin{equation}
\label{eq:TURconserv}
\eta \leq \frac{1}{1 + \frac{v\kT}{Df_{\text{ext}}}} \ \stkout{.}{\fix ,}
\end{equation}
{\fix with $D$ the motor diffusivity under nonequilibrium driving.}~\footnote{The nonequilibrium driving and track attachment break the Einstein relation between this $D$ and $\stkout{\gamma}{\fix \zeta}$.} The Stokes efficiency \stkout{(considering viscous drag, rather than pulling against a conservative force)} is similarly bounded{\fix ~\cite{pietzonka16}}, by
\begin{equation}
\label{eq:TURStokes}
\eta_{\text{Stokes}} \leq \beta \stkout{\gamma}{\fix \zeta} D \ .
\end{equation}
\stkout{Here, $\gamma$ is the friction coefficient of the motor in the medium, and $D$ is the motor diffusivity $D$ under nonequilibrium driving.}

Eqs.~\eqref{eq:efficiency1}, \eqref{eq:efficiency2}, {\fix \eqref{eq:TURconserv}, and \eqref{eq:TURStokes}} describe the efficiency of translational motors, but analogous expressions can be developed for other molecular machine types. 

With multiple molecular machines, efficiency can be enhanced, due to the many-body effects of machine interactions~\cite{Golubeva:2012bj,golubeva13}.

Living things cannot avoid spending free energy to drive molecular machines. However, resource limitations incentivize biology to reduce the free energy consumed by a particular task (all else being equal), so it is perhaps unsurprising that molecular machines can achieve remarkable efficiencies. 
$\Fo\F1$-ATP synthase has {\fix (free-energetic)} efficiency of $\sim$90\% in animal mitochondria and $\sim$65\% in chloroplasts~\cite{silverstein14}.
Although kinesin has tightly coupled ATP hydrolysis and forward steps~\cite{Schnitzer:1997bm,Hua:1997bq,Coy:1999ug}, much of the free energy kinesin consumes from ATP hydrolysis is dissipated, and its efficiency is far from unity~\cite{Ariga:2018jj}.

\subsection{Flux and output power}
\label{sec:directedMotion}
Flux $J$
measures average machine progress, or the rate at which cycles are completed.
Output power $\mcP$ is the rate at which a machine performs useful work, and is proportional to flux,
\begin{equation}
\mcP = J\Delta w \ ,
\end{equation}
where $\Delta w$ is the work per machine cycle. 
As either $J$ or $\Delta w$ is independently increased, the power increases proportionally. However, the flux itself is often a (typically decreasing) function of $\Delta w$. 
As $\Delta w$ approaches (from below) the affinity $\affin$ per cycle in the absence of a load, the flux approaches zero~\cite{Parmeggiani:1999twa,fisher99}. {\fix When $\Delta w$ reaches $\omega_{\text{tot}}$ the flux goes to zero (see \S\ref{sec:stallforce}).}

Molecular machines can achieve rapid throughput. A bacterial flagellum typically rotates $\sim$100 turns per second~\cite{Darnton:2007ct}, but can reach angular frequencies of $\sim$300 turns per second~\cite{Berg:1993bo}, as can $\Fo\F1$-ATP synthase~\cite{Ueno:2msP6jta} (exceeding jet-engine turbines)~\cite{CellBioByNum}. 
Conventional kinesin translates at 800 nm/s \emph{in vitro} (100 steps/s) and 2000 nm/s \emph{in vivo}~\cite{Howard}.
Myosin XI can reach 7 $\mu$m/s~\cite{Tominaga:2003hf}. 

Flux is important because biomolecular machines must outpace entropy increases due to the second law of thermodynamics, and act as essential players in the competition with other organisms. 
Evidence is accumulating that biological evolution values molecular machine flux, for DNA replication and tRNA selection during translation~\cite{banerjee17}, during the thermoadaptation of enzymes~\cite{nguyen17}, in the optimization of enzymes to operate near the diffusion limit~\cite{albery76}, and in bacterial adaptation machinery to sense external concentrations~\cite{lan12}. In addition to \stkout{providing organisms with faster processes} {\fix allowing organisms to increase the rate of biochemically driven processes}, fast molecular machines can conserve resources by requiring fewer machines to complete an equivalent task.

Free energy must be spent to drive directed behavior of molecular machines. While increasing free energy consumption can increase machine flux, the way in which the free energy is spent can also be adjusted to increase machine flux. Here we examine how to improve the transduction of free energy into rapid directed motion.  

Molecular machine flux is related to the driving strength. For a single-state unicyclic machine, the forward and reverse transition rate constants are related by $k^+/k^- = e^{\beta \affin}$. 
If the effect of dissipation is solely to enhance the forward rate, then the flux is $J=k^0(e^{\beta \affin} - 1)$ for bare rate constant $k^0$, giving asymptotic scaling 
$J\sim e^{\beta \affin}$ when $\beta \affin\gg 1$, and $J\sim\affin$ when $\beta \affin \ll 1$. 

For a two-stage unicyclic machine,
\begin{equation}
J = \frac{k_{12}^0 k_{21}^0 (e^{\beta \affin} - 1)}{k_{12}^0e^{\beta \omega_{12}} + k_{12}^0 + k_{21}^0e^{\beta \omega_{21}} + k_{21}^0} \ ,
\end{equation}
using Eq.~\eqref{eq:ness2states} with $k_{ij}^+ = k_{ij}^0 e^{\beta \omega_{ij}}$, $k_{ij}^- = k_{ij}^0$, and $\affin = \omega_{12} + \omega_{21}$. Although no longer straightforward, the flux depends on $\affin$, as is the case for machines with more stages or multiple cycles. 
Experiments confirm the sensitivity of molecular machine flux to affinity, specifically the dependence of kinesin flux on ATP concentration~\cite{fisher01,Lipowsky:2005ea}.

\begin{figure}[ht]
\centering
\includegraphics[width=3in]{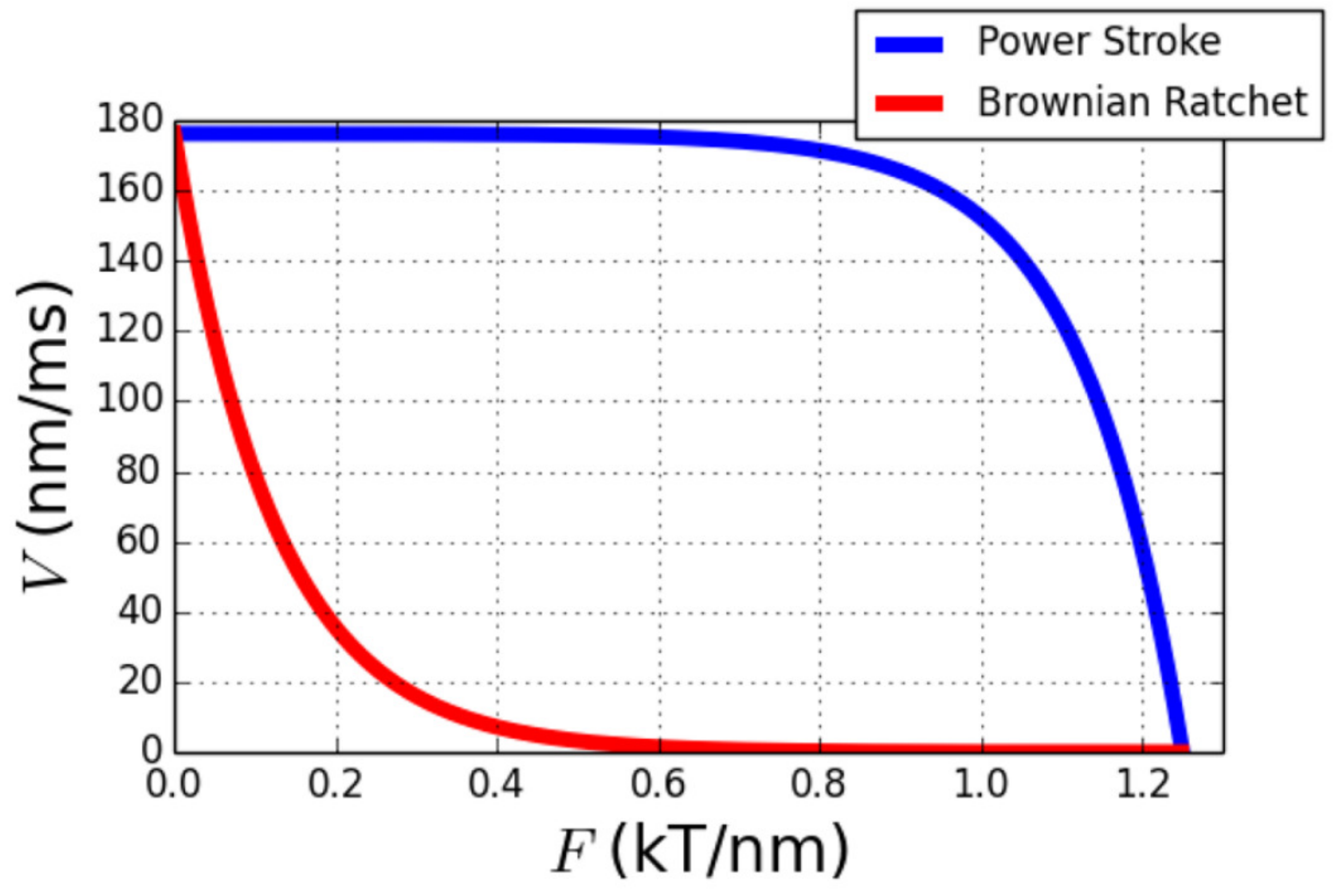}
\caption{
\label{fig:wagonerflux}
{\fix {\bf Dependence of flux on load distribution.} $F \equiv \delta w/d$ is the resisting force on a motor, and $V$ the velocity. 
When under intermediate load ($0 < F < \Delta\mu/d$), a Power Stroke (corresponding to $\delta=0$) maintains a substantially larger velocity than a Brownian Ratchet ($\delta=1$). Here, $d = 8\, \text{nm}$ and $\Delta\mu=10\, \kT$.
Reproduced from Ref.~\cite{Wagoner:2016dp}. 
Copyright 2016 American Chemical Society.}}
\end{figure}

Wagoner and Dill~\cite{Wagoner:2016dp} considered a molecular motor with a single-stage cycle and multiple distinct pathways for motor stepping. Their model included a splitting factor $\delta\in[0,1]$ {\fix (see \S\ref{sec:splittingfactor})} which divided the influence of a load 
$\Delta w$ {\fix (work done per machine cycle)} 
into the forward and reverse transition rate constants,
\begin{equation}
\label{eq:wagoner}
k^+ = k^0 e^{\beta\Delta \mu - \beta\delta \Delta w} \ \text{and} \ k^- = k^0 e^{\beta(1-\delta)\Delta w} \ . 
\end{equation}
{\fix In Eq.~\ref{eq:wagoner}, the dissipation $\beta\Delta\mu$ driving forward progress only affects the forward rate constant $k^+$ and does not change the reverse rate constant $k^-$. The influence of dissipation can generally be split among forward and reverse rate constants~\cite{Brown:2018ip}, as described by Eq.~\ref{eq:affinity2}, but for simplicity the dissipation is often assigned to only increase forward rate constants.

In contrast, in Eq.~\ref{eq:wagoner} the influence of the load $\Delta w$ is split between the forward and reverse rate constants.}
{\fix With $\delta=0$ the load only increases the reverse rate $k^-$ and leaves the forward rate $k^+$ unchanged, while for the opposite extreme ($\delta=1$) the load only slows $k^+$ and leaves $k^-$ unchanged.}
\stkout{They found that} $\delta = 0$ {\fix was found to} maximize\stkout{s} motor flux regardless of other details {\fix -- when 
optimizing
flux, it is better to accelerate reverse transitions 
than slow forward transitions. Figure~\ref{fig:wagonerflux} shows that} at zero load {\fix ($\Delta w=0$)} or stalling load {\fix ($\Delta w=\Delta\mu$)}, flux is independent of $\delta$; however, at intermediate loads a $\delta=0$ motor (where load speeds up reverse transitions) can have many times the flux of a $\delta=1$ motor (where load slows down forward transitions). Thus the distribution of a load, shared between slowing down a forward rate and speeding up a reverse rate, can {\fix substantially} influence overall motor flux.

{\fix The model of Eq.~\ref{eq:wagoner} is for a single-stage machine cycle. Equation~\ref{eq:wagoner} also, as mentioned above, splits the impact of the load $\Delta w$ between the forward and reverse rate constant, but 
models
the dissipation $\beta\Delta\mu$ 
as
only 
increasing
the forward rate constant $k^+$ while leaving the reverse rate constant $k^-$ unchanged.} Further work extended \stkout{this}{\fix the} model {\fix in Eq.~\ref{eq:wagoner}} to cycles with multiple stages and to allow splitting factors to apply to all free energy terms, not just those associated with load~\cite{Brown:2018ip}.
For this more general scenario {\fix with multiple stages and splitting of all free energy terms between forward and reverse rate constants}, there is no general optimal splitting factor $\delta$ that maximizes flux {\fix for all models.}\stkout{, and} {\fix Instead,} the optimal $\delta$ value depends on {\fix specific} details of the {\fix given} molecular machine cycle. Wagoner and Dill's result is a specific, relevant case within the more general model.

Similar to the impact of free energy division between forward and reverse transitions 
is the effect of the dissipation allocation across different stages in a multi-stage molecular machine cycle. 
The flux-maximizing allocation of free energy to the various transitions is generally uneven~\cite{brown17}. For a two-state cycle, it is
\begin{equation}
\label{eq:optimal12}
\omega_{12}^* = \frac{1}{2}\affin + \tfrac{1}{2}\kT \ln\frac{k_{21}^0}{k_{12}^0} \ ,
\end{equation}
with $\affin$ the affinity for the entire cycle, and $k_{ij}^0$ the bare rate of each transition $i\to j$. The flux-maximizing dissipation allocation depends on the bare rates $k_{ij}^0$, which describe the inherent time scale of each transition in the absence of nonequilibrium driving, combining barrier height and diffusivity along the free energy landscape. For the model leading to Eq.~\eqref{eq:optimal12}, maximal flux is achieved by allocating more dissipation to accelerate the transitions that are slower at equilibrium, essentially using nonequilibrium driving to compensate for slow equilibrium kinetics.
In the limit of large dissipation (when reverse rates are negligible), this is equivalent to equalizing the forward rates. 
The flux can depend sensitively on the dissipation allocation, decreasing by orders of magnitude only a few $\kT$ away from the optimal dissipation allocation~\cite{brown17}. The maximum flux is
\begin{equation}
\label{eq:optimalflux12}
J^* = \frac{k_{12}^0 k_{21}^0 (e^{\beta \affin} - 1)}{k_{12}^0 + k_{21}^0 + \sqrt{k_{12}^0 k_{21}^0} e^{\tfrac{1}{2}\beta \affin}} \ .
\end{equation}

\begin{figure}[ht]
\centering
\includegraphics[width=3.5in]{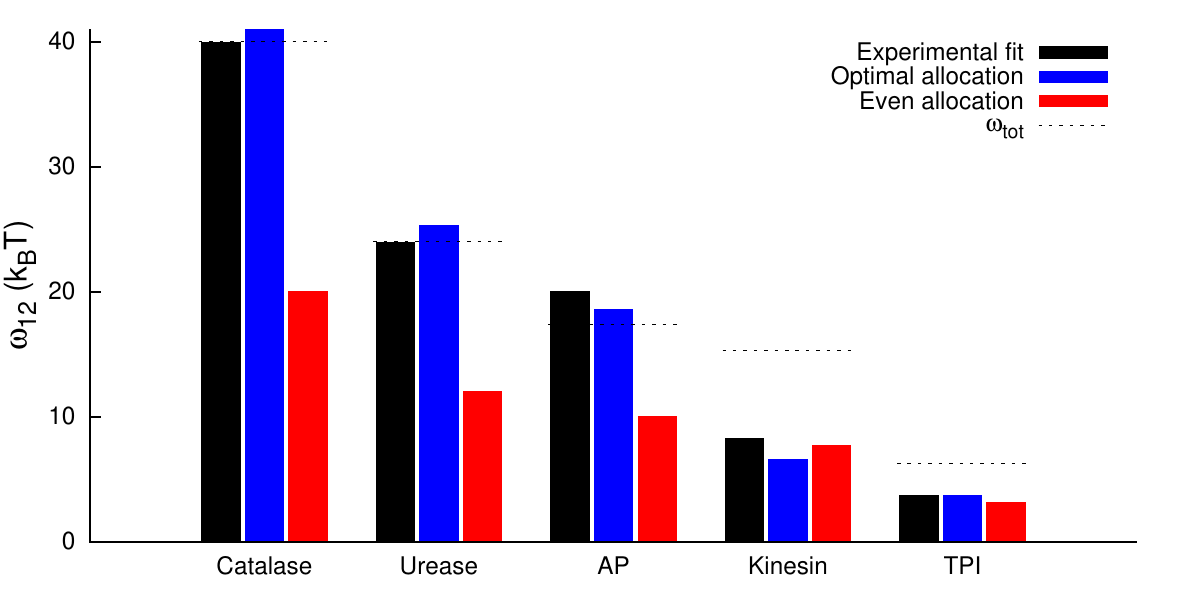}
\caption{
\label{fig:PNASfig6}
{\bf Maximum flux predictions for dissipation allocation.}
For several enzymes, comparison of dissipation allocation between fit to experiment (black), allocation that maximizes flux from Eq.~\eqref{eq:optimal12} (blue), and even allocation (red). $\omega_{12}$ is the larger dissipation in a two-state discrete model, and $\omega_{\text{tot}} = \omega_{12}+\omega_{21}$.
Experimental fits from Hwang and Hyeon~\cite{hwang17}. AP, alkaline phosphatase; TPI, triose phosphate isomerase. 
Adapted from Ref.~\cite{brown17}.}
\end{figure}

Figure~\ref{fig:PNASfig6} shows two-state models for several machines, parameterized from experimental data. The dissipations fit to experiment are well-described by uneven allocation~\eqref{eq:optimal12}, while in three cases, even allocations are clearly inconsistent with the experimental fits.

\begin{figure}[ht]
\centering
\includegraphics[width=3.5in]{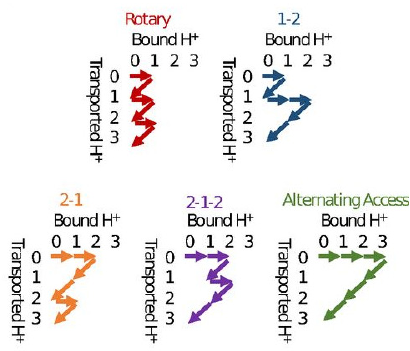}
\caption{
\label{fig:anandakrishnan}
{\bf Various schemes for proton-transport order of ATP-synthase-like enzymes.}
Possible mechanisms for ATP-synthase-like enzymes based on proton transport order for a 3:1 $\text{H}^+$:ATP stoichiometry (`Rotary' labels actual mechanism). Addition and removal of bound H$^+$ correspond to increases and decreases of the free energy landscape. 
Adapted from Ref.~\cite{Anandakrishnan:2016jz}.}
\end{figure}

Anandakrishnan et al.~\cite{Anandakrishnan:2016jz} explored free energy use by a molecular machine that requires several unfavorable
transitions.
This scenario is inspired by ATP synthase operation, which over one rotation binds three ADP molecules and releases them as three ATP molecules. The process of catalyzing each ADP$\to$ATP reaction requires the binding (and hence the driving force) of three protons, involving both a free energy increase (upon binding a proton) and a later free energy decrease (upon releasing a proton) -- how should these free energy increases and decreases be ordered to maximize ATP synthase flux (Fig.~\ref{fig:anandakrishnan}) ? 
Mechanistically, it may seem simpler for a machine like ATP synthase to bind three protons, phosphorylate ADP to ATP, and release ATP and the three protons. 
Anandakrishnan found that simultaneously processing three ATP and sequentially binding and releasing individual protons (corresponding to repeated but modest free energy increases followed by decreases) results in a faster cycle than all alternative reaction schemes. The rotary mechanism of ATP synthase allows this repetitive cycle, avoiding unnecessarily high free energy increases that would decrease the cycle flux, by dividing what could be a single high free energy barrier into several lower barriers.

Wagoner and Dill~\cite{Wagoner:2019gm} more recently further explored how characteristics of free energy use control molecular machine flux, bringing existing and new results into a single framework. They found that equal barrier heights maximize flux, consistent with earlier work~\cite{brown17}. Using free energy from fuel to cancel out free energy costs of mechanical work also leads to higher flux, similar to aligning multiple free energy components such that together they are optimal~\cite{Brown:2018ip}. 
For a wider range of models, they showed that an early transition state ($\delta = 0$) increases machine flux. Echoing the results of Anandakrishnan et al.~\cite{Anandakrishnan:2016jz} for ATP synthase, Wagoner and Dill found that higher flux is achieved by
splitting up the cost of external work across many stages.

Overall, recent work has shown that the details of a free energy landscape (dissipation allocation, ordering of increases/decreases, and splitting of free energy for each transition) can be adjusted to speed up molecular machine operation, even if the affinity, or free energy budget, remains unchanged. It is important to note that 
the optimal free energy landscape depends on the applied constraints, so there is no universally optimal choice~\cite{brown17,Brown:2018ip}.

\subsection{Precision}
\label{sec:precision}
The stochastic nature of molecular machine operation 
inextricably produces {\fix substantial} variation in the behavior of an individual machine. 
The precision of an individual machine's motion is commonly defined as the \emph{Fano factor} of the number $n$ of machine cycles completed,
\begin{equation}
\label{eq:fano1}
F \equiv \frac{\langle n^2\rangle-\langle n\rangle^2}{\langle n\rangle} \ .
\end{equation}
The Fano factor $F$ is defined as the ratio between the variance and mean of a stochastic quantity{\fix~\cite{Fano:1947}}.
For a transport motor such as kinesin, $n$ is the net number of forward steps taken.
\footnote{{\fix Quantifying precision highlights that a molecular machine could serve as a `clock'~\cite{Barato:2016fo}, and indeed the distinction between a given system functioning as motor or clock may be fuzzy~\cite{Hess:2017clock}.}}

For a transport motor taking steps of size $d$, the average progress is the average distance $\langle X\rangle=d\langle n\rangle$, and the Fano factor is
\begin{equation}
{\fix F} = \frac{\langle X^2\rangle - \langle X\rangle^2}{d\langle X\rangle} \ .
\end{equation}
Substituting the diffusivity $D \equiv (\langle X^2\rangle - \langle X\rangle^2)/(2t)$ and mean velocity $v \equiv \langle X\rangle/t$ reformulates the Fano factor as the \emph{randomness parameter}~\cite{svoboda94,fisher01},
\begin{equation}
\label{eq:randomnessparameter}
r \equiv \frac{2D}{vd} \ .
\end{equation}
Machines with small step-number Fano factors or randomness parameters have more reliable behavior and are generally considered to be higher performing.
The randomness parameter has been directly measured in kinesin experiments, across a wide range of forces and ATP concentrations, to be $0.3\lessapprox r\lessapprox1.5$~\cite{Schnitzer:1997bm,Visscher:1999fe,block03}.
\stkout{\S 
discusses precision in greater detail in the context of the thermodynamic uncertainty relation.} 

\stkout{The stochastic operation of molecular machines leads to substantial variability in the progress of individual molecular machines. 
In \S\ref{sec:precision}, we defined the Fano factor of the net number of machine cycles completed~\eqref{eq:fano1}, 
which is closely related to the randomness parameter~\eqref{eq:randomnessparameter}.}

For a general unicyclic motor with $N$ states, the Fano factor has a lower bound~\cite{koza02,koza02b},
\begin{equation}
\label{eq:randomnessnumber}
F \geq \frac{1}{N} \ .
\end{equation}
Motors with more states per cycle can achieve greater precision, provided the transitions have similar rate constants~\cite{svoboda94}. Equation~\eqref{eq:randomnessnumber} has been used to estimate the number of states in the cycles of biomolecular motors~\cite{block03}.

Barato and Seifert derived a tighter limit on motor precision that also incorporates free energy consumption~\cite{barato15}. 
For a single-stage (i.e.\ one-state) motor with forward rate constant $k^+$ and reverse rate constant $k^-$, the mean net number of forward transitions or 
steps is $\langle n\rangle = (k^+ - k^-)t$ and the variance is $\langle n^2\rangle-\langle n\rangle^2 = (k^+ + k^-)t$. The rate-constant ratio is related to the affinity, $k^+/k^- = \exp[\beta \affin]$ (special case of Eq.~\eqref{eq:affinity2}). 
Thus the dissipation rate is $\Omega = \affin\langle n\rangle/t = (k^+ - k^-)\affin$.
Combining the dissipation rate with uncertainty $\epsilon^2 \equiv (\langle n^2\rangle-\langle n\rangle^2)/\langle n\rangle^2 = (k^+ + k^-)/[(k^+ - k^-)^2t]$, yields the \emph{thermodynamic uncertainty relation}
\begin{equation}
\label{eq:bsfano1}
\Omega t\epsilon^2 = \affin\coth\tfrac{1}{2}\beta\affin \geq 2\kT \ .
\end{equation}
Equation~\eqref{eq:bsfano1} relates affinity $\affin$ to achievable uncertainty $\epsilon^2$. For example, precision of 1$\%$, or $\epsilon^2=10^{-4}$, requires $\affin \geq 2\times10^4 \kT$.

For an $N$-state motor cycle, Barato and Seifert additionally showed that the precision of motor progress is limited by~\cite{Barato:2015bt}
\begin{equation}
\label{eq:bsfano}
\frac{\langle n^2\rangle - \langle n\rangle^2}{\langle n\rangle} \geq \frac{1}{N}\coth\frac{\beta\affin}{2N} \geq \frac{1}{N} \ .
\end{equation}
Equation~\eqref{eq:bsfano} generalizes Eq.~\eqref{eq:randomnessnumber} to finite dissipation, as Eq.~\eqref{eq:randomnessnumber} represents the $\affin\to\infty$ limit of Eq.~\eqref{eq:bsfano}. Equation~\eqref{eq:bsfano} quantifies the greater precision that autonomous motors can achieve with a larger number $N$ 
of discrete states per cycle and/or larger affinity $\affin$.

Barato and Seifert initially proved Eq.~\eqref{eq:bsfano1} for unicyclic networks and conjectured that it would hold more generally~\cite{barato15,Barato:2015bt}; their conjecture was soon proven by Gingrich et al. using large deviation theory~\cite{gingrich16}. The thermodynamic uncertainty relation~\eqref{eq:bsfano1} has been generalized to finite times~\cite{pietzonka17,horowitz17} and discrete time~\cite{proesmans17}, and the bound can tighten for special cases~\cite{polettini16}.
Conversely, simply adding a large (and hence slowly diffusing) towed cargo to a molecular motor{\fix~\cite{Zimmermann:2012es,Zimmermann:2015hh,Zimmermann:2015ub}} can considerably increase {\fix the} \stkout{this} achievable precision, because the effective number $N$ of motor states is sensitive to strong coupling of the motor to its surroundings~\cite{brown18cargo}. 

The thermodynamic uncertainty relation~\eqref{eq:bsfano1} can be applied widely, constraining heat engines in addition to isothermal motors~\cite{pietzonka18}, and can be used to constrain other performance metrics, such as the efficiency of a molecular motor pulling against a constant force~\cite{pietzonka16}. Beyond motors and machines, the thermodynamic uncertainty relation has been used to constrain molecular copying~\cite{ouldridge17} and self-assembly processes~\cite{nguyen16}. 


Assessment of several cases shows biomolecular motor operation approaches the thermodynamic uncertainty relation limit~\cite{hwang18}. Equality in Eq.~\eqref{eq:bsfano1} can be achieved by {\fix a} process\stkout{es} with {\fix a} Gaussian distributions of dissipation, \stkout{both near and} {\fix arbitrarily} far from equilibrium~\cite{hyeon17}.

\subsection{Stall force}
\label{sec:stallforce}
Molecular motors can work against both conservative external forces and viscous forces that oppose the motion of the motors themselves or their towed cargo~\cite{qian04}.
When the forces opposing a motor reach the \emph{stall force}, the mean motor velocity drops to zero. 
Machines with higher stall forces are able to make forward progress in a broader range of conditions.

Stall force is limited by the free energy available per motor cycle with zero load~\cite{qian00} (the affinity $\affin$ at zero load) and the motor step size $d$~\cite{fisher99}:
\begin{equation}
\label{eq:stallforce}
f_{\text{stall}} \leq \frac{\affin}{d} \ .
\end{equation}
Equality in Eq.~\eqref{eq:stallforce} corresponds to a \emph{tightly coupled} motor, where consumption of each unit of fuel leads to a forward step~\cite{Bustamante:2004fo}. 
In contrast, a \emph{loosely coupled} motor has backsteps that consume free energy, or futile cycles where free energy is consumed but no steps are taken, thereby decreasing the stall force~\cite{Bustamante:2004fo}. 

Of course, higher affinity $\affin$ produces a higher stall force $f_{\text{stall}}$. Less obviously, longer individual steps $d$ lead to a lower stall force for a given affinity. 

Kinesin takes 8 nm steps and stalls at $\sim$4 pN (in the presence of 2 
mM ATP)~\cite{clancy11}, compared to myosin V that takes 36 nm steps and stalls at 1.6 pN (in the presence of 2 mM ATP)~\cite{cappello07}, and dynein that takes 8 nm steps and stalls at anywhere from 1-7 pN depending on species~\cite{Nicholas:2015bz}. RNA polymerase can achieve a stall force as high as 30 pN~\cite{Wang:1998vc}.

\subsection{Processivity}
Molecular motors can take many consecutive steps along their intracellular filament tracks before eventually detaching. The number of such steps taken before detachment is a stochastic quantity, with the typical number defining the \emph{processivity}. As the primary role of many of these molecular motors is to transport cargo along filaments, more processive motors (that take more steps before detachment) are considered to have better performance.

Transport motors can achieve impressive processivity.
For example, conventional kinesin typically takes $\sim$100 steps before detaching~\cite{Schnitzer:1997bm}. Myosin V has a characteristic run length of $\sim$1400nm (corresponding to $\sim$40 36-nm steps) under small load~\cite{sakamoto04}.

\subsection{Specificity}
Certain {\fix bio}molecular machines \stkout{that} copy a sequence of chemical letters or {\fix transcribe/}translate one sequence type into another: \stkout{(such as DNA polymerase, RNA polymerase, and ribosomes)} {\fix DNA polymerase copies a DNA sequence, RNA polymerase transcribes the DNA sequence into an mRNA sequence, and} ribosomes translate a sequence of RNA codons (groups of three nucleotides) into the corresponding sequence of amino acids. {\fix These machines} have `soft' parts and thus must rely on small energy differences, comparable to $\kT$, to discriminate between correct and incorrect copies. 
This contrasts notably with macroscopic machines, which can be designed with macroscopic energy penalties that are prohibitively large (e.g.\ a component that does not fit). 
{\fix Thus} these {\fix (microscopic)} biomolecular copying and translation processes have a nonzero error rate. 

Despite these features, molecular machines can achieve low error rate (high \emph{specificity}), quantified by the fraction of letters in the product that incorrectly represents the initial template~\cite{AlonSysBio}. 
The error rate for DNA replication has been measured as $10^{-10}-10^{-8}$, for RNA transcription as $10^{-5}-10^{-4}$, and for protein translation as $10^{-4}-10^{-3}$~\cite{banerjee17}.
\stkout{\S\ref{sec:trade-offs} further discusses specificity in the context of kinetic proofreading.} 

\subsection{Performance trade-offs}
\label{sec:trade-offs}
\stkout{Many measures of performance, such as those outlined in \S\ref{sec:performance}, can be used to evaluate molecular machines.} 
Variation of operational parameters can alter these measures of performance. One might imagine that optimizing various measures of performance would lead to ideal molecular machine operation; however, a given molecular machine design cannot simultaneously optimize all measures of performance. 

\begin{figure}[ht]
\centering
\includegraphics[width=3.5in]{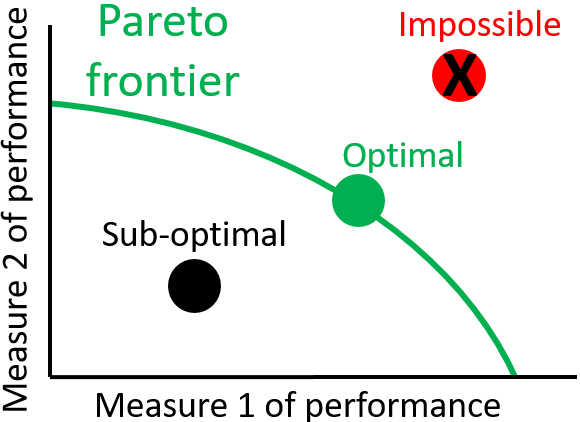}
\caption{
\label{fig:pareto}
{\bf The Pareto frontier} represents a range of possibilities for optimal performance, given the unavoidable trade-offs between distinct measures of performance. Points not on the Pareto frontier (e.g.\ labeled `Sup-optimal') are worse on at least one performance measure than a point on the Pareto frontier (e.g.\ labeled `Optimal'), and no better at any other tasks. Points beyond the Pareto frontier (e.g.\ labeled `Impossible') are not achievable.}
\end{figure}

The Pareto frontier~\cite{shoval12,sheftel13} (Fig.~\ref{fig:pareto}) is defined by the set of points in the space of performance measures for which improving one performance measure requires degrading other distinct measures, leading to trade-offs between them. Essentially, the Pareto frontier constitutes the set of achievable combinations of performance on various measures, such that there is no unambiguously better achievable combination.  

Multi-objective optimization, 
where the relative importance of different performance measures is varied,
can be used to identify the Pareto frontier. 
The frontier can be non-convex, implying a phase transition (rather than a continuous crossover) in optimal strategy as the relative importances are varied~\cite{Solon:2018gv}.

In the sections above, we have already outlined some trade-offs. Given the focus of this review on how dissipation relates to molecular machine operation, these trade-offs include flux vs.\ dissipation~\eqref{eq:optimalflux12}, precision vs.\ dissipation (Eqs.~\eqref{eq:bsfano1} and \eqref{eq:bsfano}), and stall force vs.\ dissipation~\eqref{eq:stallforce}. 

\begin{figure}[ht]
\centering
\includegraphics[width=3.5in]{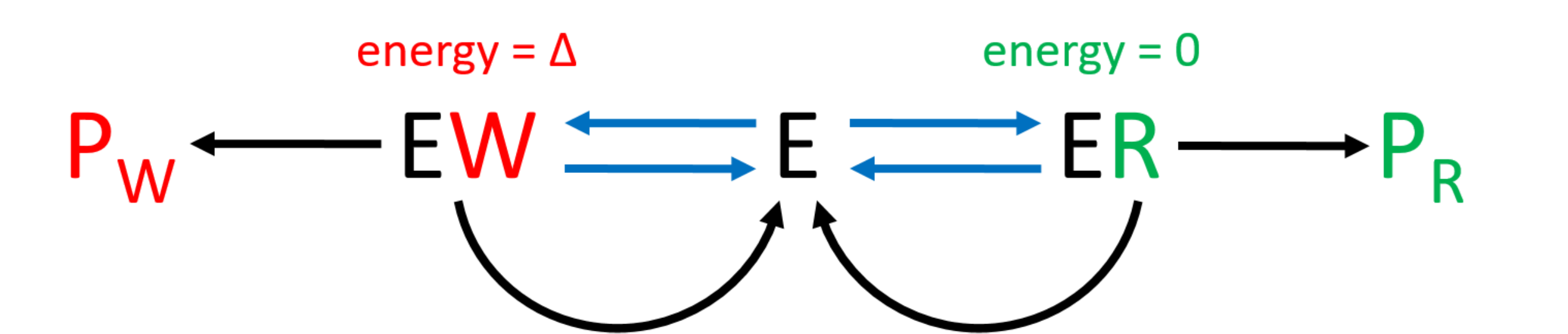}
\caption{
\label{fig:proofreading}
{\bf Simplified kinetic proofreading} scheme similar to those in DNA replication, RNA transcription, tRNA aminoacylation (charging), and protein translation. Straight blue arrows in two directions represent reversible binding of the right substrate (R, green) or the wrong substrate (W, red) to a copying enzyme (E), and straight black arrows the permanent incorporation of the substrate into the copying product (P). Curved black arrows represent cyclic sampling of many candidate substrates for the copy, driven by the dissipation of free energy. {\fix The two arrow types from ER and EW to E represent distinct physical processes: the straight blue arrows represent reversal of the binding process, while the curved black arrows represent the removal of R and W using free energy dissipation and are not simple reversals of the respective binding processes.} Combined with energy penalty $\Delta$ for wrong substrates, cyclic sampling substantially decreases the probability of selecting the wrong substrate.
}
\end{figure}

\stkout{Another measure of performance is the specificity of copying or translation by biomolecular machines. DNA and RNA polymerase make copies of or transcribe nucleotide sequences, with the individual nucleotides in the sequence often known as `letters.'}  
Here we discuss {\fix trade-offs between dissipation and specificity of copying/transcribing} sequences of DNA/RNA {\fix nucleotides} (letters) \stkout{and replication/transcription}; similar principles apply to amino acids and translation.
The copying enzyme has `right' (R) and `wrong' (W) substrates competing to be added as the next letter in the sequence (Fig.~\ref{fig:proofreading}). The enzyme goes through intermediate stages before permanently incorporating R into the right product $\rm P_{\rm R}$ or W into the wrong product $\rm P_{\rm W}$. For a wrong substrate with an energy penalty $\Delta$ 
relative to the right substrate, an equilibrium process \stkout{will have} {\em has} an error rate $e^{-\beta \Delta}$. However, this copying process is driven out of equilibrium by free energy consumption to cyclically sample (represented by cycles in Fig.~\ref{fig:proofreading}) many possible right and wrong substrates, which can decrease the error rate to a minimum bound of $e^{-2\beta \Delta}$~\cite{hopfield74,ninio75,murugan12}. This relationship represents a trade-off between specificity and free energy consumption, and decreasing the error rate from the equilibrium value to the nonequilibrium value depends on the free energy used per cycle to drive the process out of equilibrium.

For given energy differences and nonequilibrium driving, there is also a trade-off between copying speed 
and specificity. Lower error rates come at the cost of many sampling cycles, which decrease copying speed. Recent work has explored this trade-off, finding that DNA copying and protein translation are near the Pareto frontier for speed and \stkout{accuracy} {\fix specificity}~\cite{banerjee17}. This work also found that DNA copying and protein translation \emph{in vivo} appear to favor speed over {\fix specificity} \stkout{accuracy} once a certain level of {\fix specificity} \stkout{accuracy} is reached.

Beyond even a broad definition of molecular machines, other \stkout{systems} {\fix processes} such as T-cell activation~\cite{mckeithan95,francois13}, bacterial chemotaxis~\cite{lan12,sartori15}, and signal transduction~\cite{Mehta:2012ji} also demonstrate trade-offs between energy, speed, and accuracy.

\section{External driving costs}
\label{sec:deterministicDriving}
Free energy transduction requires input of free energy. A common framework is to guide the machine by varying a control parameter, driving the machine between distinct macrostates, and imparting energy to the machine. Such externally imposed control of a molecular machine comes at the thermodynamic cost of dissipated energy which is no longer available to perform work. Given the tasks, models, and physical principles underlying molecular machine operation described above, we now examine the energy cost for deterministically driven molecular machine dynamics.

In single-molecule experiments, such explicitly time-dependent driving can be imposed externally by an experimentalist; \emph{in vivo} operation of autonomous molecular machines is not subject to such externally imposed dynamical protocols. 
Rather, such driving can come from another mechanically coupled molecular component, often another machine or another part of the given machine (e.g. in ATP synthase, $\Fo$ mechanically rotating the central crankshaft to drive $\F1$). 
In this sense, an upstream machine component can mimic the nonequilibrium variation provided by an experimentalist.

\subsection{Directional differences}
Directed behavior of molecular machines can derive from asymmetric response to external perturbations~\cite{Hoffmann:2016gb}. The distinctiveness of the system response can be quantified by the time asymmetry $A$, which quantifies the ease of distinguishing between the forward and reverse trajectory distributions for an externally driven process. 
This can be information-theoretically expressed as the Jensen-Shannon divergence $\text{JS}$ of the forward and reverse trajectory distributions~\cite{feng08},
\begin{equation}
\label{eq:asymmetryjs}
A[\Lambda] \equiv \text{JS}(P[x|\Lambda],\tilde{P}[\tilde{x}|\tilde{\Lambda}]) \ ,
\end{equation}
{\fix where}
\begin{equation}
{\fix \text{JS}(p,q) \equiv \frac{1}{2}\sum_i \left[p_i\ln\frac{p_i}{\frac{1}{2}(p_i + q_i)} + q_i \ln\frac{q_i}{\frac{1}{2}(p_i + q_i)}\right] \ ,}
\end{equation}
is a measure of the distance between two probability distributions $p$ and $q$~\cite{Cover:2006:Book}.
Here, $P[x|\Lambda]$ is the probability of forward trajectory $x$ during forward protocol $\Lambda$\stkout{, and $\tilde{P}[\tilde{x}|\tilde{\Lambda}]$ is its time-reversed counterpart}. {\fix $\tilde{x}$ and $\tilde{\Lambda}$ are the time-reverse of trajectory $x$ and protocol $\Lambda$, respectively, such that $\tilde{P}[\tilde{x}|\tilde{\Lambda}]$ is the probability of a reversed trajectory when the protocol is run backwards. Time asymmetry $A$ represents the expected distinguishability of system trajectories 
when
driven by the forward protocol $\Lambda$ 
compared to 
its time-reversed counterpart $\tilde{\Lambda}$.}

The Jensen-Shannon divergence contains two Kullback-Leibler divergences, 
\begin{equation}
{\fix D(p,q) \equiv \sum_i p_i \ln\frac{p_i}{q_i} \ ,}
\end{equation}
which have {\fix also} been used to quantify irreversibility~\cite{lacasa12,roldan12}. In contrast to the Kullback-Leibler divergence, the Jensen-Shannon divergence is symmetric with respect to the two input probability distributions, such that switching the `forward' and `reverse' assignments does not change its value.

The Jensen-Shannon divergence of the trajectory distributions~\eqref{eq:asymmetryjs} can be rewritten~\cite{Crooks:2011:JStatMech} using the Crooks theorem~\eqref{eq:crooks} in terms of work distributions from driving in both directions~\cite{feng08}:
\begin{equation}
\label{eq:asymmetryw}
A[\Lambda] = \frac{1}{2}\left\langle\ln\frac{2}{1 + \exp(-\beta W[x|\Lambda] + \beta\Delta F)}\right\rangle_{\Lambda} + \frac{1}{2}\left\langle\ln\frac{2}{1 + \exp(-\beta W[\tilde{x}|\tilde{\Lambda}] - \beta\Delta F)}\right\rangle_{\tilde{\Lambda}} \ ,
\end{equation}
where $W[x|\Lambda]$ is the work done over system trajectory $x$ during forward protocol $\Lambda$, {\fix and} $\Delta F$ is the equilibrium free energy difference over the forward protocol. This makes time asymmetry experimentally measurable.

{\fix A trajectory ensemble under protocol $\tilde{\Lambda}$ that is not the time reversal of the trajectory ensemble under $\Lambda$ indicates a departure from equilibrium and requires free energy dissipation. This free energy} \stkout{The} cost is {\fix quantified by} the hysteretic dissipation
\begin{equation}
\label{eq:asymmetrydissipation}
h[\Lambda] \equiv \frac{1}{2}\left(\beta \langle W[x|\Lambda] \rangle + \beta \langle W[\tilde{x}|\tilde{\Lambda}] \rangle\right) \ ,
\end{equation}
the average work for the forward and reverse protocols, equal to the average excess work because the free energy differences cancel~\cite{feng08}. 

{\fix Low time asymmetry indicates that a system responds similarly to driving in both directions, while high time asymmetry indicates distinct response 
depending on the driving direction. This distinct response to driving could facilitate the directed dynamics required for molecular machine function, and the time asymmetry generated for a given amount of free energy dissipation 
assesses
how effectively the free energy has been spent for the purpose of generating directed behavior. Accordingly,} biological evolution may 
\stkout{be expected to} 
favor biomolecular machines that achieve a high time asymmetry for a given dissipation cost, 
{\fix motivating investigation into} \stkout{and thus we ask} 
what process characteristics lead to relatively high asymmetry~\eqref{eq:asymmetryw} for a given dissipation~\eqref{eq:asymmetrydissipation}.

\begin{figure}[ht]
\centering
\begin{tabular}{cc}
\includegraphics[height=2in]{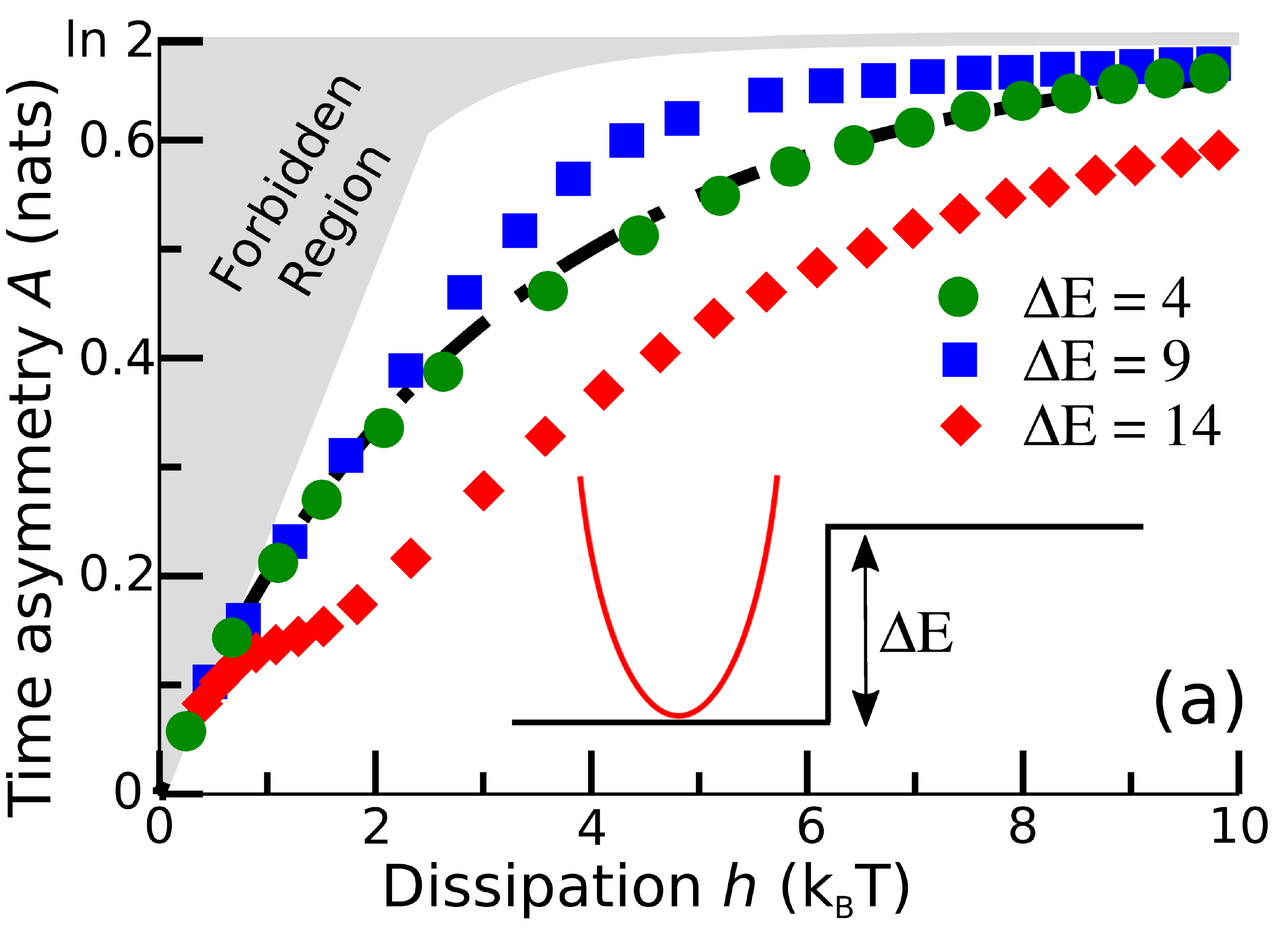}&
\includegraphics[height=2in]{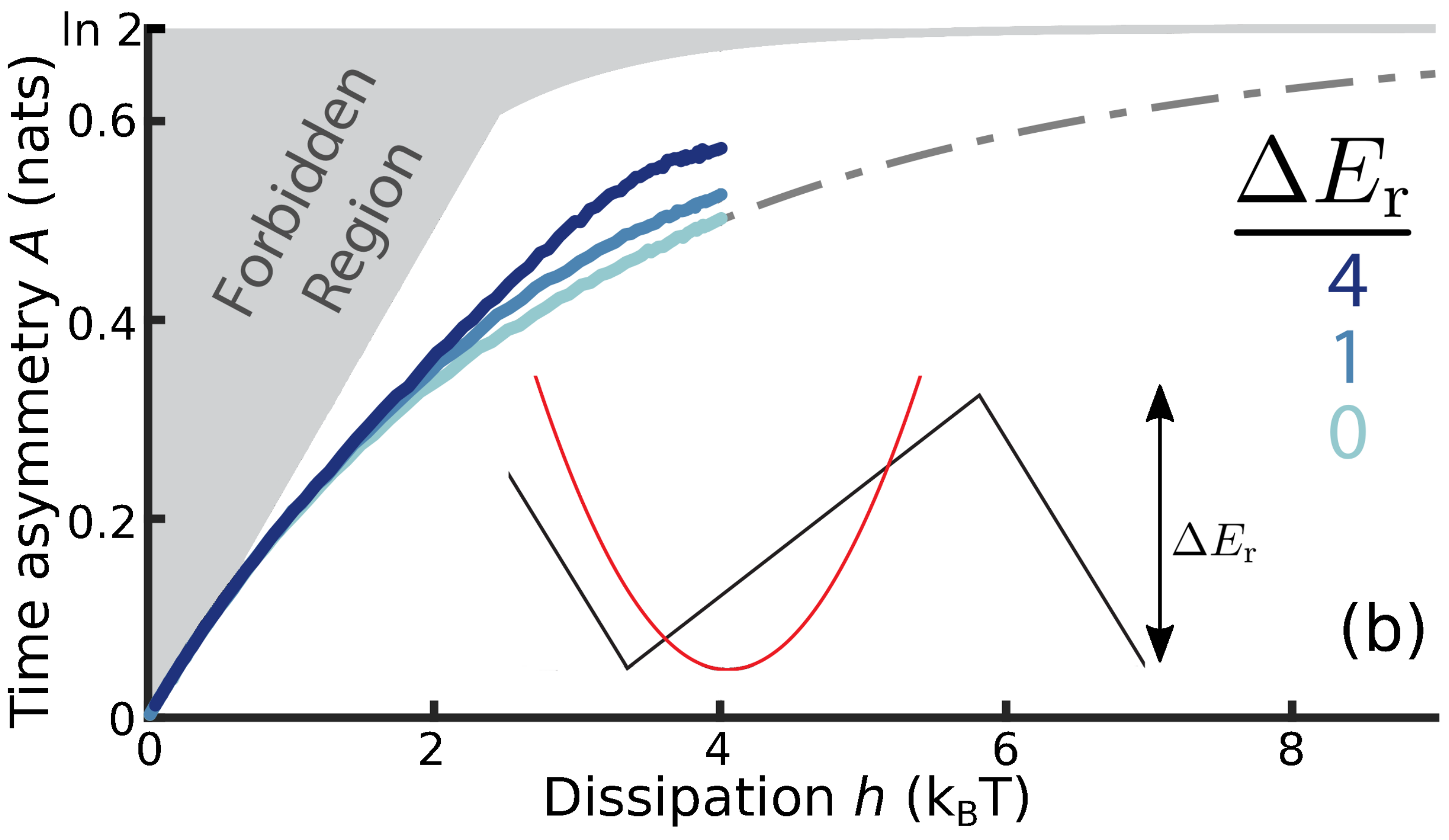}
\end{tabular}
\caption{
\label{fig:timeasymmetry}
{\bf Time asymmetry} $A$ vs.\ dissipation $h$. 
(a) Green circles, blue squares, and red diamonds: parametric plots 
\stkout{of time asymmetry vs.\ dissipation} when driving a system up a step potential 
for low, intermediate, and high steps, respectively.
{\fix (b) Light, medium, and dark blue points: parametric plots for driving a system across a sawtooth for negligible, low, and higher sawteeth.}
Different points for each color represent different driving speeds, with higher time asymmetry $A$ and dissipation $h$ from faster driving. 
\stkout{Solid} {\fix Dot-dashed} black curve: linear response.
Gray region: unfeasible time asymmetries for a given dissipation. 
Adapted with permission from (a) Ref.~\cite{brown16} and (b) Ref.~\cite{zarrin18}. Copyright 2016 and 2019 American Physical Society, respectively.}
\end{figure}

{\fix Initial} empirical investigations, including experiments~\cite{collin05,feng08} and molecular dynamics simulations~\cite{procacci10}, found time asymmetry trade-offs with dissipation that remain near the linear-response prediction~\cite{feng08}. 
{\fix In contrast, using a step-function energy landscape (Fig.~\ref{fig:timeasymmetry}a, inset) to represent an idealized} \stkout{For a} free energy storage process (with $\Delta F>0$), intermediate \stkout{free energy changes} {\fix step heights} 
lead to time asymmetries that can exceed linear response for a given dissipation~\cite{brown16} (Fig.~\ref{fig:timeasymmetry}{\fix a}). 
{\fix Additionally, using a sawtooth energy landscape (Fig.~\ref{fig:timeasymmetry}b, inset)} \stkout{In contrast, for protocols} with $\Delta F = 0$ {\fix (frequently used to represent molecular machines)}, high and symmetric intervening barriers {\fix between consecutive states} lead to high asymmetry for a relatively low dissipative cost~\cite{zarrin18} (Fig.~\ref{fig:timeasymmetry}{\fix b}). 

{\fix Both step-function and sawtooth energy landscapes allow time asymmetry to exceed linear response, although with distinct strategies -- intermediate-height step function 
contrasting with
high sawtooth.} In both cases, {\fix the system is driven by translating a quadratic energy trap, with medium-strength traps maximizing time asymmetry by balancing} reliable driving to a given final state {\fix{(achieved with a narrow trap at high energy cost)} \stkout{competes} against the increas{\fix ed}\stkout{ing} dissipative cost associated with more tightly constraining a system}\stkout{, such that systems constrained to medium-sized fluctuations produce higher time asymmetries}~\cite{brown16,zarrin18}. 
Overall, driving a system such that it can transiently `get stuck' can provide distinct forward and reverse trajectory distributions that lead to higher time asymmetries for a given dissipation~\cite{zarrin18}.

Beyond finite transitions driven by control protocols, the relationship between thermodynamic costs and reliable directed behavior of stochastic processes can be extended to NESS processes. The entropy production per time $\md S/\md t$ to generate a NESS with directed behavior in a finite average time $\langle \tau \rangle$ is
\begin{equation}
\frac{1}{k_{\text{B}}}\left\langle\frac{\md S}{\md t}\right\rangle = \frac{L(1-2\alpha)}{\langle \tau\rangle} \ ,
\end{equation}
where $L$ is the threshold amount of progress for directed behavior (how far in the forward or reverse direction the system must progress), and $\alpha$ is the probability that the system will run in the correct direction~\cite{roldan15}. Faster processes, and those with a higher threshold for sufficient progress, have a higher entropy cost per unit time for directed behavior.

\subsection{Deterministic driving}
\label{sec:friction}
When a control parameter $\lambda$ changes sufficiently slowly, a system will remain near equilibrium and be well-described by linear-response theory~\cite{Zwanzig:2001,Chandler1987a}. For a system adhering to linear response, it can be shown that the instantaneous excess power (i.e.\ beyond that required when the system is equilibrated throughout the protocol) required during a control protocol (\S\ref{sec:protocols}), due to the system being out of equilibrium, is~\cite{sivak12} 
\begin{equation}
\label{eq:power}
\mcP_{\rm ex}(t_0) = \zeta(\lambda(t_0))\cdot\left(\frac{\md\lambda}{\md t}\right)^2_{t=t_0}\ .
\end{equation}
$\zeta$ is a generalized friction coefficient that governs near-equilibrium response and represents the energetic cost of changing the control parameter sufficiently fast to drive the system out of equilibrium:
\begin{equation}
\label{eq:friction}
\zeta(\lambda(t_0)) \equiv \beta\int_0^{\infty} \md t\langle\delta f(0)\delta f(t)\rangle_{\lambda(t_0)} \ .
\end{equation}
Here $\delta f \equiv f - \langle f\rangle_{\lambda(t_0)}$ is the deviation of the conjugate force $f$ from its equilibrium average $\langle f\rangle_{\lambda(t_0)}$ at fixed control parameter value $\lambda(t_0)$. 
For instance, when the control parameter is the separation between two optical traps~\cite{Tafoya:2019hr}, the conjugate force is the tensile force with which the biomolecule resists further extension. 
$\langle\delta f(0)\delta f(t)\rangle_{\lambda(t_0)}$ is an autocorrelation function: at $t=0$ it equals the force variance {\fix $\langle \delta f^2 \rangle_{\lambda(t_0)}$}, while for $t>0$ \stkout{the autocorrelation function} {\fix it} represents how quickly the system forgets its initial condition.
{\fix The generalized friction $\zeta(\lambda)$ can be decomposed into the product of the force variance and the integral relaxation time
$\tau(\lambda) \equiv \int_0^{\infty} \md t \frac{\langle \delta f(0) \delta f(t)\rangle_{\lambda}}{\langle \delta f^2\rangle_{\lambda}}$, the characteristic lifetime of force fluctuations.}

Equation~\eqref{eq:friction} is an example of a fluctuation-dissipation theorem relating equilibrium fluctuations to dissipation out of equilibrium. 
In particular, $\zeta$ is a Green-Kubo coefficient~\cite{Kubo:1957vy} expressing a transport coefficient as a temporal integral of a correlation function, in this case the generalized friction coefficient in the space of control parameters as an integral of the force autocorrelation function.

How do we minimize the excess work, the time integral of \eqref{eq:power} over a control protocol of fixed duration, to waste as little energy as possible? 
In general, proceeding slowly reduces the excess work, such that in the quasistatic limit the excess work is zero, with the work equaling the equilibrium free energy difference. 
Equation~\eqref{eq:power} expresses the 
rate of accumulation of excess work along a control protocol.
For protocols that minimize work in a fixed duration, the excess power is constant, which is achieved when the control parameter is varied according to~\cite{sivak12}: \begin{equation}
\label{eq:optimal}
\frac{\md\lambda^{\rm opt}}{\md t} \propto \frac{1}{\sqrt{\zeta(\lambda(t))}} \ .
\end{equation}

Implementation of this theory only depends on being able to measure, at fixed control parameter, fluctuations of the force conjugate to the control parameter. This gives a phenomenological procedure, which does not rely on detailed knowledge of either system kinetics or thermodynamics. (If one does have such detailed knowledge, then protocols that minimize work far from equilibrium can be found~\cite{Schmiedl:2007:PhysRevLett,GomezMarin:2008:JChemPhys,Aurell:2011wz}.)

Several groups have used this framework to examine optimal protocols in model systems~\cite{sivak12,zulkowski12,Zulkowski:2014:PhysRevE,Bonanca:2014:JChemPhys,Rotskoff:2015:PhysRevE,Zulkowski:2015:PhysRevE:b,Rotskoff:2017:PhysRevE,Bonanca:2018dm}. 
Applying this theory to bistable systems representing thermally activated processes~\cite{sivak16} leads to the intuition that energetically efficient control requires relatively slow perturbation when the system is on the verge of a major transition, essentially letting random thermal fluctuations kick the system over a given barrier `for free' without energy input from the controller.
Other extensions have generalized this control framework to nonequilibrium steady states~\cite{Mandal:2016:JStatMech,zulkowski13} and to models of rotary machines~\cite{Lucero:2019gd} and chemical reaction networks~\cite{Blaber:2018}. 

\begin{figure}[ht]
\centering
\includegraphics[width=\textwidth]{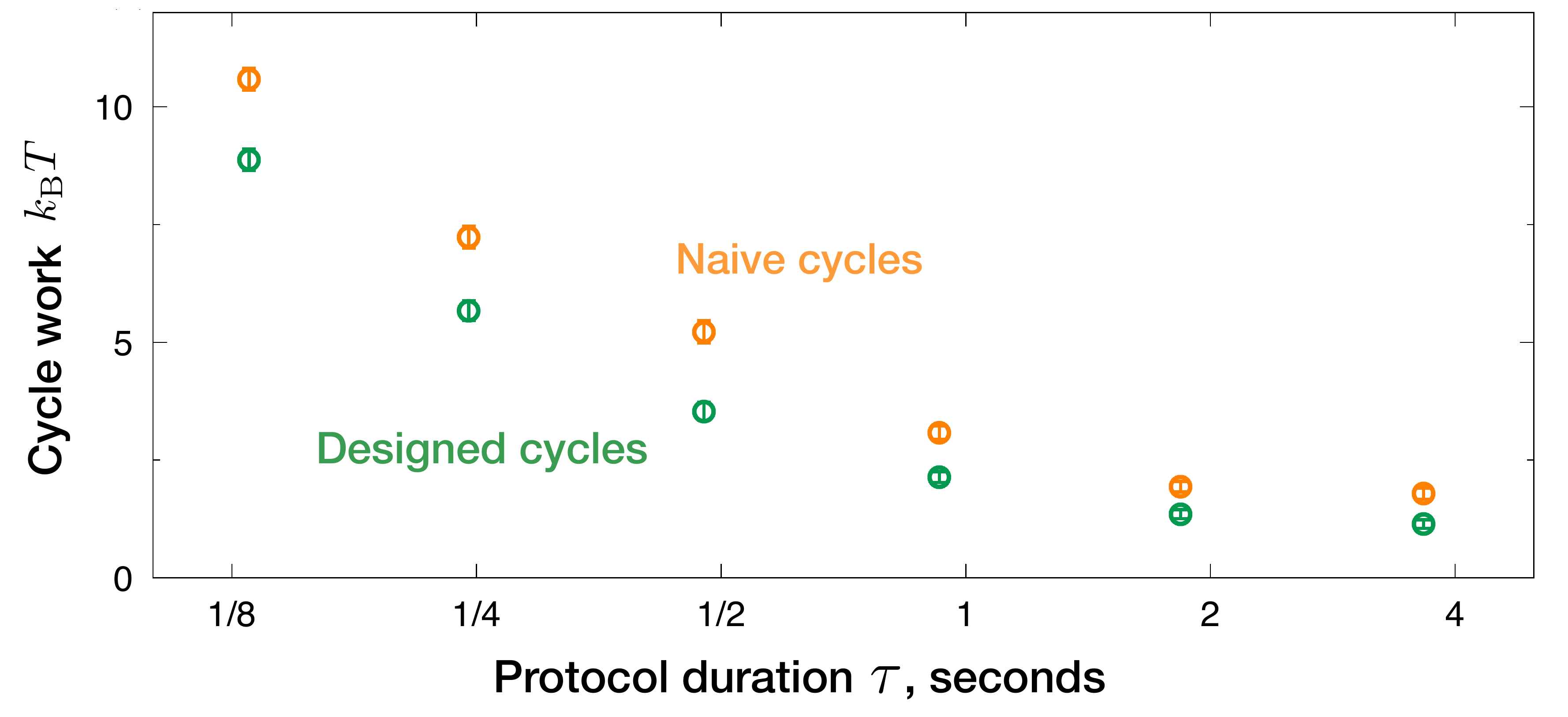}
\caption{
\label{fig:hairpin}
{\bf Designed protocols to unfold and refold DNA hairpins systematically require significantly less work than corresponding naive protocols.} 
Mean cycle work $\langle W^{\rm Unfolding} + W^{\rm Refolding}\rangle$ during naive (green) and designed (orange) protocols as a function of protocol duration.
Reproduced from Ref.~\cite{Tafoya:2019hr}.}
\end{figure}

Proof-of-principle experiments recently demonstrated the utility of this theory for predicting energy-conserving behavior in nanoscale biophysical systems~\cite{Tafoya:2019hr}, using a DNA hairpin (a small piece of DNA that spontaneously folds up on itself) as a model system. 
Measurements of force fluctuations at fixed optical-trap separation identified that the generalized friction is maximized in the `hopping' regime where the DNA hairpin is equally likely to be folded or unfolded. 
The optical traps were then dynamically modulated to rapidly stretch the ends of the DNA hairpin (putting work into the hairpin and unfolding it), and then bring the hairpin ends close together (allowing the hairpin to refold and thereby recovering work). 
The energy lost during cycles designed to proceed slowly through the hopping regime was significantly less than during ‘naive’ cycles that proceeded at a constant speed (Fig.~\ref{fig:hairpin}). 
This energy savings was found systematically at a variety of cycle speeds and for two different hairpins differing dramatically in their relaxation time.

\subsection{Stochastic driving}
\label{sec:stochasticDriving}
The theory in \S\ref{sec:friction} focuses on deterministic driving of a system, where a control parameter follows a fixed temporal schedule. Biomolecular machines do not typically experience an experimentalist deterministically changing a control parameter -- instead they operate autonomously, responding to the stochastic fluctuations of coupled nonequilibrium systems. 

For example, in ATP synthase, the $\F1$ subunit is driven by another subunit, $\Fo$, that itself operates stochastically. 
Experimental observation of $\F1$ rotational statistics indicates a small number of metastable angular states separated by energetic barriers~\cite{Yasuda:1998:Cell}.
When the angle of a magnetic trap (the control parameter for single-molecule driving of $\F1$, see \S\ref{sec:caseStudies}) is centered at a barrier separating two adjacent metastable states, equilibrium probability is split equally between the two states, giving maximal torque variance $\langle \delta f^2 \rangle$ and maximal torque relaxation time, hence maximizing their product, the friction coefficient $\zeta$~\cite{Lucero:2019gd}.

Equation~\eqref{eq:optimal} provides intuition on how an experimentalist (or $\Fo$ \emph{in vivo}) should drive rotation to minimize energy expenditure: where the friction coefficient is large---where the system puts up large resistance to rapid control parameter changes, at the rotational energetic barriers---the minimum-dissipation protocol proceeds slowly, giving thermal fluctuations maximal time to kick the system over the barrier `for free.' 

Since $\Fo$ itself is stochastic, and hence cannot impose deterministic driving protocols on $\F1$, it would have to resort to a stochastic mimic of such a designed protocol. In particular, if $\Fo$ itself has metastable rotational states out of phase with those of $\F1$, then the stochastic protocol would amount to rapid rotations followed by pauses at rotational states corresponding to the hopping regime where $\F1$ at equilibrium is evenly split between two rotational states, thereby forming a near-analog of the designed deterministic protocols discussed above. 

Evolved machines provide tantalizing hints of out-of-equilibrium behavior that would reduce energy consumption. 
For example, the $\phi$29 DNA packaging motor 
slows down (packaging shorter stretches of DNA with longer intervening pauses) as the $\phi$29 viral capsid is increasingly filled~\cite{Liu:2014:Cell,Berndsen:2014jf}.  
This behavior is consistent with predictions of the linear-response control theory, as the relaxation time for the DNA in the capsid increases strongly with packing fraction. 
Similarly, translating ribosomes appear to `change gear' when encountering an RNA hairpin that impedes translation: the ribosome slows down while surmounting the energetic barrier represented by the hairpin~\cite{Chen:2015gt}.
Both examples could be interpreted as driving protocols that proceed slower where the friction coefficient is higher, thereby predicted to reduce dissipation. 

\begin{figure}[ht]
\centering
\includegraphics[width=\textwidth]{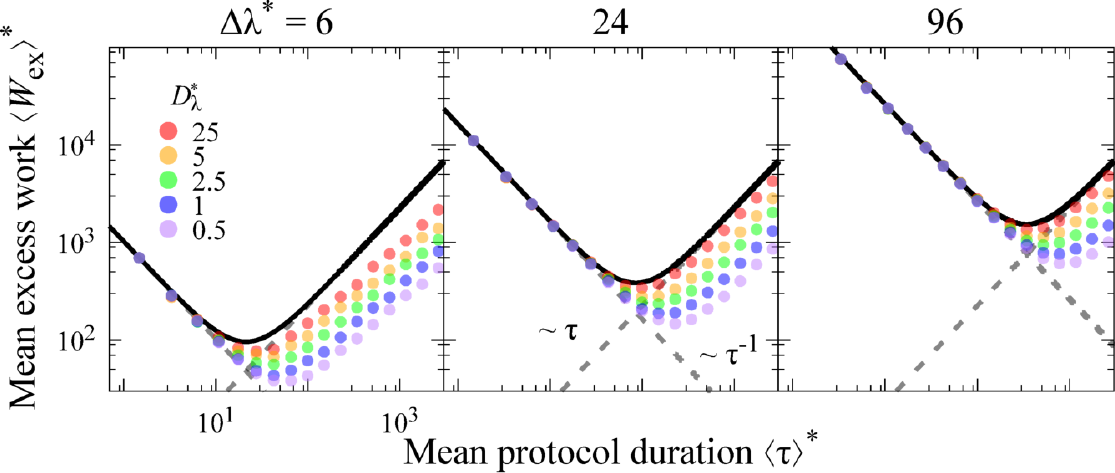}
\caption{
\label{fig:stochastic_control}
{\bf Excess work for a stochastic protocol is minimized at finite protocol duration.}
Mean nondimensionalized excess work $\langle W_{\rm ex}\rangle^* \equiv \beta \langle W_{\rm ex}\rangle$ as function of mean protocol duration $\langle \tau \rangle^* \equiv \tau/(2\beta D k)^{-1}$.  For underdamped control parameter dynamics and large nondimensionalized protocol distances $\Delta \lambda^* \equiv \Delta \lambda/(\beta k)^{-1}$ (for harmonic trap spring constant $k$)---where control parameter dynamics are locally deterministic---numerical simulations (circles) agree with the linear-response approximation (solid black curve), the sum of terms proportional and inversely proportional to protocol duration (dashed black curves). 
Control parameter diffusion coefficient $D^*_{\lambda}\equiv D_{\lambda}/D$ (nondimensionalized by the system diffusion coefficient) interpolates between overdamped (purple) and underdamped (red) control parameter dynamics. 
Reproduced with permission from Ref.~\cite{Large:2018dh}. 
Copyright 2018 European Physical Society.} 
\end{figure}

Motivated by stochastic mechanical driving in molecular machines, the study of such stochastic protocols---where the control protocol does not evolve deterministically, but rather with its own stochastic dynamics---\stkout{leads to} {\fix reveals} an interesting qualitative feature: 
{\fix stochastic control-parameter fluctuations on average require net work, which continually accumulates along the protocol. Thus}
\stkout{because} 
work due to stochastic fluctuations increases with protocol duration;  \stkout{(whereas for deterministic protocols longer durations require less work), for stochastic protocols work is minimized} 
{\fix since the work due to the mean protocol decreases with duration, the resulting trade-off leads to stochastic protocols minimizing work} at intermediate protocol durations~\cite{Large:2018dh} (see Fig.~\ref{fig:stochastic_control}).

Extension of these ideas to discrete driving protocols (inspired by the effectively instantaneous nature of chemical reactions, compared to other relevant time scales) reveals a thermodynamic cost that remains even in the quasi-static limit~\cite{Large:Discrete}.

Active efforts currently seek to situate the developing framework of stochastic and discrete control parameter protocols within the more fully developed stochastic thermodynamics of chemical reaction networks~\cite{Rao:2016ga,Rao:2018kr,Rao:2018ij}.

\subsection{Autonomous driving}
\S\ref{sec:friction} and \S\ref{sec:stochasticDriving} focused on the dissipation due to system resistance to deterministic or stochastic control protocols. 
But the very implementation of a control parameter whose dynamics break detailed balance requires dissipation, independent of the dissipation associated with system resistance to the control protocol. When a single-molecule experimentalist implements control of a molecular machine, this cost of control may not be a limiting factor {\fix . However,}  \stkout{; but} for autonomous machines, notably \emph{in vivo} molecular machines, there is a cost associated with implementing particular asymmetric driving dynamics, which should be included in any accounting of total dissipative costs. 

Recent research~\cite{machta15,Verley:2014hi}
on these `upstream' thermodynamic costs associated with imposing particular time-asymmetric, detailed-balance breaking protocol ensembles~\cite{machta15,BryantMachta} has opened new vistas on the physical principles governing autonomous molecular machines driven by stochastic protocols. 
Current research is examining the connections between these imposition costs and the `downstream' dissipation resulting from system resistance to the stochastic control parameter dynamics~\cite{Large:2018dh}. 
Interesting questions also remain about how such thermodynamic flows within strongly coupled systems relate to work and heat flows between weakly coupled systems and the environment~\cite{Verley:2014hi}. 




\section{Synthetic molecular machines} \label{sec:synthetic}
Although extensive discussion is beyond the scope of this review, many of the ideas we discuss here are being explored from an engineering perspective in synthetic machines~\cite{erbas-cakmak15}. 
Development of synthetic molecular machines includes pumps, walkers, transporters, and rotary motors~\cite{wilson16,DelRosso:2017kb,Wang:2019fh}. In addition to helping us better understand evolved machines, synthetic machines are of interest in their own right because of technological applications{\fix ~\cite{Hess:2011:AnnRevBiomedEng}, including} in computation and information storage/retrieval (memory)~\cite{Zulkowski:2014:PhysRevE}, renewable energy (artificial photosynthesis)~\cite{McConnell2014}, drug delivery, and other biomedical goals~\cite{Peng2017}. 

Synthetic molecular machines have been built or designed that are driven by light~\cite{balzani06}, electricity~\cite{tierney11}, cycled chemical concentrations~\cite{zuckermann15,small19}, or consuming their track~\cite{kovacic15,korosec18}, among other driving modes~\cite{erbas-cakmak15}. Systems driven by cycling chemical concentrations will not reach a NESS due to \stkout{variable transition} {\fix time-varying} rate constants, although constant and cyclic driving are equivalent in some other respects~\cite{raz16}. 
In \S\ref{sec:microscopicreversibility}, transitions were described as obeying microscopic reversibility. 
For some systems driven by light, microscopic reversibility will not be followed, {\fix as these transitions are not driven by thermal fluctuations over the energy landscape.}
Instead{\fix, these transitions involving light} follow\stkout{ing} the Einstein relations for absorption and stimulated and spontaneous emission~\cite{astumian15}. 

Although most synthetic molecular machines are driven in an altogether different manner than biomolecular machines, an autonomous synthetic molecular machine was recently developed that is driven by nonequilibrium chemical concentrations. The machine involves a small ring transported around a cyclic molecular track with a preferential direction~\cite{wilson16}.
It remains to be seen whether molecular machines that operate similarly to their biomolecular counterparts will gain wider use.


\section{Emerging ideas}
\label{sec:emerging}
In this section we outline some emerging concepts in machine transduction.

\subsection{Information machines}
In most of this review, we discuss how free energy is used to drive molecular machine operation. Here we briefly outline an alternative `fuel' for driving molecular machines: information.

The second law of thermodynamics dictates that the entropy of an isolated system remains constant or increases. Therefore, in a container of equilibrated gas the fast (hot) molecules do not spontaneously segregate to one side with the slow (cold) molecules on the other side. However, if an intelligent agent was able to selectively open and close a door separating the two sides, only allowing slow molecules to leave one side, and fast molecules to leave the other side, then the temperature of one compartment would increase while the temperature of the other compartment decreases: heat would flow from cold to hot, thereby decreasing entropy. This scenario was originally proposed by Maxwell~\cite{maxwell71}, with the intelligent agent known as `Maxwell's demon'~\cite{Maruyama:2009gh,Bennett:2002th}. 
Szilard's engine~\cite{szilard29} is a similar scenario where knowledge of a gas molecule's location is used to extract work from the thermal bath. If the intelligent agent can operate without free energy consumption, both Maxwell's demon and Szilard's engine appear to violate the second law.

Information has a physical manifestation~\cite{parrondo15},
famously expressed in Landauer's principle, which says that information erasure has a minimum free energy cost of $\kT\ln 2$ per bit~\cite{landauer61}, which has been confirmed experimentally~\cite{Berut:2012jd,jun14}. The intelligent agents of Maxwell's demon and Szilard's engine must pay this information erasure cost, saving the second law~\cite{bennett82}.

\begin{figure}[ht]
\centering
\includegraphics[width=3in]{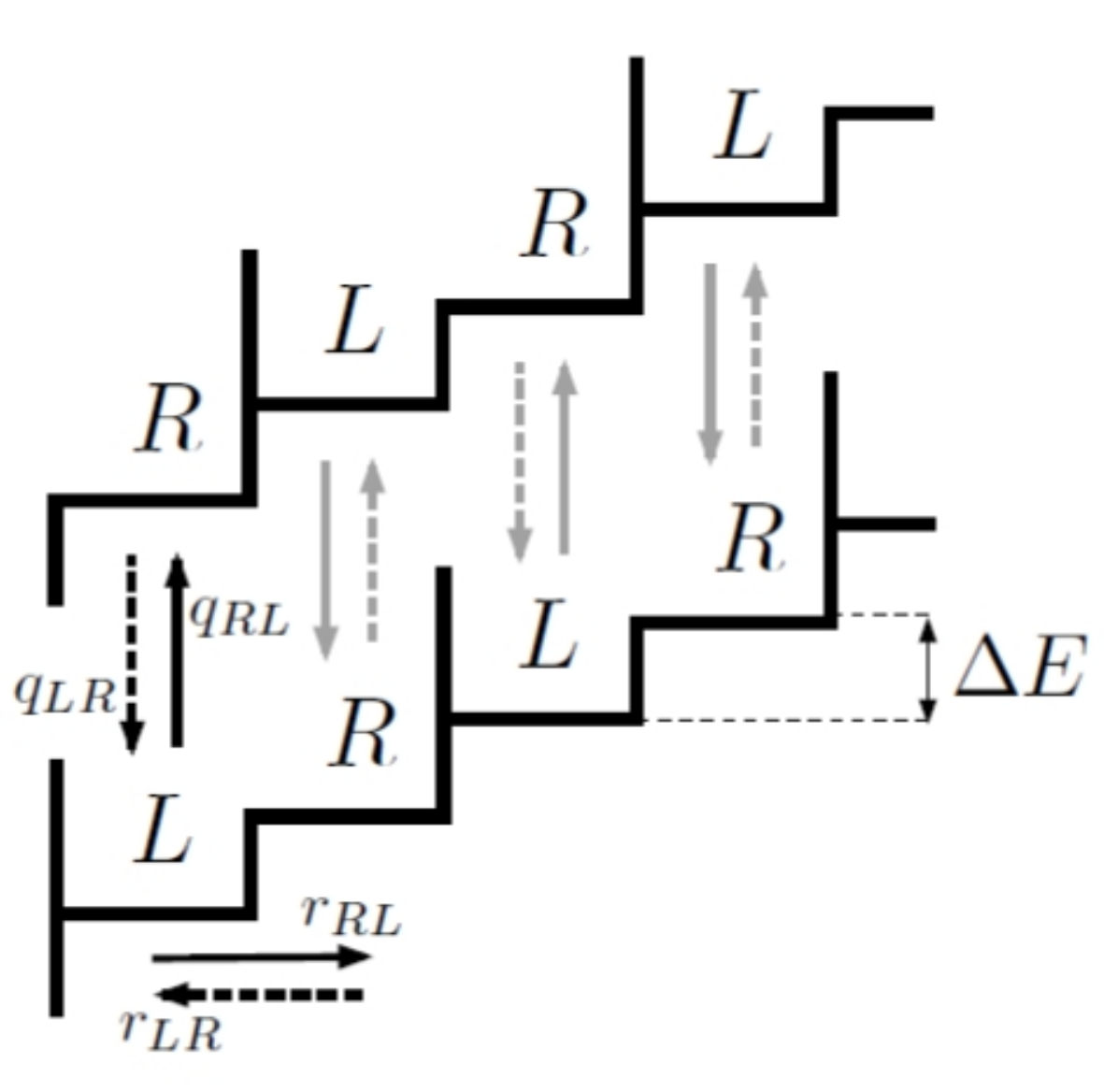}
\caption{
\label{fig:infoMotor}
{\bf Information motor.} A system moves in a periodic potential, with two energy levels, with the higher energy level in the direction of desired progress. The potential switches between two configurations, changing which neighboring energy level is accessible from each energy level. If information about the particle location is used to preferentially switch the barriers, the system can progress uphill. 
Adapted with permission from Ref.~\cite{horowitz13}. Copyright (2013) by the American Physical Society.}
\end{figure}

Molecular machines can make forward progress and do useful work using feedback. One implementation~\cite{horowitz13,Toyabe:2010dn} involves a staircase of states separated by a constant energy interval (Fig.~\ref{fig:infoMotor}). Infinite barriers separate each pair of states from its neighbors on both sides. The system is thus confined to a pair of states, with the forward state at a higher energy. The barrier locations can shift to adjust the accessible pair of states, depending on the system position\stkout{:}{\fix.} If the system is in the higher-energy {\fix of the accessible pair of} states when the shift occurs, {\fix the previously accessible lower-energy neighboring state is no longer accessible, and instead the system} \stkout{it} can now access the next higher{\fix-energy} state up the staircase\stkout{;}{\fix.} If the system is in the lower-energy state when the shift occurs, {\fix the previously accessible higher-energy state is no longer accessible, and the system} \stkout{it} can now access the next lower state down the staircase. 
If the barriers can be preferentially switched when the system is in the higher state, the system energy increases.
Such preferential switching involves feedback and memory, and such information processing has a corresponding \stkout{equivalent} free energy cost~\cite{horowitz13}. Another implementation used feedback to `push' a particle upstream against flow without doing work~\cite{Admon:2018by}.

Including information processing leads to modified forms of the second law~\cite{sagawa09} that can account for the classic thermodynamic problems of Maxwell's demon and Szilard's engine, along with the newer ideas of information machines. In addition to providing a novel driving mode, information machines have distinct thermodynamic behavior, such as linear-response efficiency~\cite{barato14} or free energy dissipation~\cite{horowitz13}.

\subsection{Predictive machines}
\S\ref{sec:deterministicDriving} focused on controlling a machine, rapidly driving it to a new desired macrostate. By contrast, here we take the perspective of the machine, discussing how a system can couple effectively to its environment. 
{\fix\S\ref{sec:performance} generally} examined machine dynamics that perform well given average nonequilibrium aspects of the environment; here we discuss utilizing fluctuations in a nonequilibrium environment, specifically how properties of a system and its interactions with its environment affect transduction of energetic fluctuations into increased free energy, and hence capacity for work.

The difference between the average work $\langle W\rangle$ imparted to the system and the resultant change in the system's capacity for useful work, its nonequilibrium free energy $F_{\rm neq} \equiv \langle E\rangle - TS$~\cite{Hasegawa:2010dm,Sivak:2012:PhysRevLett:a}, is the dissipated work $W_{\rm diss} = \langle W\rangle - {\fix \Delta} F_{\text{neq}}$.  
{\fix (Note that for quasistatic processes---when environmental changes are sufficiently slow that the system remains at equilibrium throughout---the nonequilibrium free energy reduces to the equilibrium free energy $F_{\rm eq} = \langle E\rangle_{\rm eq} - TS_{\rm eq}$, and the dissipated work vanishes.)} 
The current system state $s_t$ retains information about the current environmental state $x_t$, quantified by the mutual information $I_{{\rm mem}}(t) \equiv I[s_t; x_t]$~\cite{Cover:2006:Book}. Mutual information $I$ quantifies how much (on average) knowledge of the current system state reduces uncertainty about the current environmental state. Because the system is coupled to its environment, and in general an environment does not immediately randomize its state, the current system also has predictive information about the future environmental state $x_{t+1}$, quantified by $I_{{\rm pred}}(t) \equiv I[s_t; x_{t+1}]$. The dissipated work during environmental dynamics equals the difference between these two informations, the unpredictive information (or \emph{nostalgia}) the system retains about the environment, which does not predict the future state of the environment~\cite{still12}:
\begin{equation}
\label{eq:memory}
\beta\langle W_{\rm diss}[x_t\to x_{t+1}]\rangle = I_{\rm mem}(t) - I_{\rm pred}(t) \ .
\end{equation}

\stkout{Thus} {\fix This equivalence of thermodynamic inefficiency $\beta\langle W_{\rm diss}\rangle$ and predictive inefficiency $I_{\rm mem}-I_{\rm pred}$ means that} any system, be it an organism, a neuron, or even a single molecular machine, that has been designed, over natural evolution or in a nanoscientist's lab, to be thermodynamically efficient in its interactions with its environment, must (at least implicitly) be constructing a parsimonious {\fix (not overly complex)} model of environmental dynamics~\cite{still12,Quenneville:2018ig}. 
At steady state this nostalgia 
(for unit-time steps)
equals the learning rate, the rate at which a system (due to its own dynamics) increases its mutual information with the environment~\cite{Hartich:2014bv,Barato:2014fg,Hartich:2016gs,Barato:2013jx,Brittain:2017hf}.

Intuitively, this says that thermodynamic efficiency is accomplished by forgetting (i.e., rapidly randomizing and hence relaxing to equilibrium with respect to) degrees of freedom in one's environment that are not predictive of future fluctuations.
Possible implications for efficient transduction of environmental fluctuations by molecular machines are that they must ignore the myriad aspects of their environment (e.g., sundry collisions from water molecules) that have no bearing on mechanically meaningful future environmental fluctuations. 
Conversely, temporal correlations in environmental perturbations can provide thermodynamic advantage if a system can learn them.

\subsection{Dissipative adaptation}
This review has emphasized \stkout{physical} limits and design principles for effective utilization and/or avoidance of dissipation, because high-performance molecular machines should be favored by natural selection. 
Recent research further suggests a relationship between free energy dissipation and the evolution of systems in the absence of natural selection.

England and co-workers have shown within the framework of stochastic thermodynamics~\cite{Seifert:2012es} that there is a relationship between the degree of irreversibility and required entropy production for macroscopic transitions~\cite{england13},
\begin{equation}
\label{eq:england1}
\beta \langle \Delta Q\rangle_{\mathbf{I}\to\mathbf{II}} + \ln \frac{\pi(\mathbf{II}\to\mathbf{I})}{\pi(\mathbf{I}\to\mathbf{II})} + \frac{\Delta S_{\text{int}}}{k_{\text{B}}} = 0 \ ,
\end{equation}
where $\langle\Delta Q\rangle_{\mathbf{I}\to\mathbf{II}}$ is the heat dissipated as the system transitions from state $\mathbf{I}$ to $\mathbf{II}$, $\pi(\mathbf{I}\to\mathbf{II})$ is the probability of a system initially at state $\mathbf{I}$ transitioning to state $\mathbf{II}$, and $\Delta S_{\text{int}}$ is the entropy change of the system between $\mathbf{I}$ and $\mathbf{II}$.
However, England also argues that this relationship between irreversibility and entropy production has consequences for self-replicating processes~\cite{england13}:   
\begin{equation}
\label{eq:england2}
g_{\text{max}} = \chi e^{\beta \Delta q + \Delta s_{\text{int}}/k_{\text{B}}} \ . 
\end{equation}
Here, $g_{\text{max}}$ is the maximum growth rate, $\chi$ is the decay rate, and $\Delta q$ and $\Delta s_{\text{int}}$ are intensive versions of the earlier quantities. Equation~\eqref{eq:england2} argues that the maximum growth rate is limited by the dissipation, durability, and organization: higher dissipation allows higher growth rates, and increased durability and organization decrease the allowed growth rate.

It follows that systems are more likely to follow trajectories that have higher dissipation. 
Therefore a systems can `adapt' (in the absence of natural selection) to dissipate free energy of the form that is available, because those dissipative trajectories are more likely~\cite{england15}. If a system is self-replicating, this adaptation can be taken further, as systems which follow trajectories allowing improved reception of the provided free energy will haven successors that continue to follow more likely trajectories~\cite{perunov16}. 

\subsection{Enzyme diffusion enhancement} 
\label{sec:enhancedDiffusion}
Energy can be consumed to create and enhance directed motion, but recent studies have found that energy consumption can also enhance random undirected motion (diffusivity). 
Essentially, free energy can be transduced into quicker motion or dynamical spreading. 
Specifically, carrying out catalysis has been found to increase the diffusive mobility of enzymes~\cite{yu09,muddana10}. This diffusion enhancement can be substantial, with the largest observed enhancement of $\sim$80\%~\cite{xu18}. This effect increases with higher substrate concentrations, and appears limited to enzymes performing catalysis, as it is not shared with nearby enzymes~\cite{sengupta13}. Although higher diffusion occurs at higher temperatures, overall catalytic energy release does not appear to be the cause, as diffusion enhancement has been measured for both endothermic and exothermic reactions~\cite{illien17}. 
Global energy changes and changes to charge and pH have also been excluded as mechanisms~\cite{riedel15}. 
Enhancement of enzyme diffusion was initially measured with fluorescence correlation spectroscopy (FCS). Recently, the possibility was raised that artifacts in FCS measurements could lead to an apparent increase in measured diffusion without an actual increase in diffusion~\cite{gunther18}, but enhanced enzyme diffusion has been confirmed with direct single-molecule imaging~\cite{xu18}.

Efforts to systematically relate enzyme diffusion enhancement to other measurable quantities have met with some success. For some exothermic reactions the diffusivity enhancement is proportional to the catalyzed reaction rate and the heat released by the reaction, with an enzyme-specific proportionality constant~\cite{riedel15}. 
The increased enzyme velocity is thought to be sustained for only a very short period of time, on the order of nanoseconds, following each catalyzed reaction~\cite{riedel15}. It has been suggested that enzymes with low proportionality constants between reaction rate and released heat (i.e.\ enzymes whose diffusion is not significantly enhanced), e.g.\ catalase, may be able to achieve higher throughput because the low diffusivity enhancement may indicate relative conformational stability~\cite{riedel15}. 

There is no consensus for the underlying mechanism driving enhanced enzyme diffusion during catalysis. It has been proposed that the enhanced enzyme diffusivity is due to a `chemoacoustic' effect, where the asymmetric expansion and contraction of the enzyme during and following catalysis generates heat and an asymmetric pressure wave which pushes back on the enzyme to generate the enhanced diffusion~\cite{riedel15}. Alternatively, differences in fluctuation ensembles between substrate-bound and -unbound enzyme states could contribute to enhanced diffusivity, particularly for enzymes with low throughput that catalyze endothermic reactions~\cite{illien17}.

Enhanced enzyme diffusion may have biological implications, possibly allowing enzymes to chemotax (move to higher substrate concentrations) and draw together chemically connected enzymes~\cite{sengupta13}. Beyond just isolated enzymes, a similar effect would reposition the complex of nucleic-acid copying enzymes and their template (i.e.\ DNA/RNA polymerase and DNA) to higher nucleotide concentrations~\cite{yu09,sengupta14}. Interestingly, it has also been suggested that enhanced enzyme diffusion could allow enzymes to antichemotax (move to lower substrate concentrations), which could smooth out the spatial distribution of enzymatic production~\cite{jee18b}.

Recent work has argued that enzyme diffusion enhancement due to catalysis fits within the same framework that describes the diffusion enhancement of traditional motors, such as kinesin, due to dissipation~\cite{hwang17}.


\section{Conclusions}
\label{sec:conclusion}
By writing that ``living matter evades the decay to equilibrium,'' Schr{\"o}dinger pointed out the inextricable link between free energy consumption and life~\cite{schrodinger44}. 
Molecular machines, as pivotal free energy transducers in all living systems, are major participants in this life-sustaining dissipation. 

In this review, we (almost certainly incompletely) summarize recent theoretical efforts in several areas involving free energy transduction by molecular machines.
As theorists, we are excited about recent and newly emerging theoretical frameworks to describe and understand molecular machines, and we are continually astounded and thankful for the ongoing experimental innovation that is absolutely essential for continued advance of this field.


Theory, perhaps especially in biophysics, faces a constant struggle between generalizing to an abstract and all-encompassing framework, while meaningfully engaging with a messy and heterogeneous reality.
Recognizing this tension, we provisionally suggest a potentially unifying principle cutting across many molecular machines: effective transduction between different nonequilibrium free energy reservoirs. More broadly, we emphasize the importance of identifying and engaging with a set of stylized facts that together form a consensus description for the operation of many molecular machines. 

Nevertheless, despite extensive quantitative characterization of a handful of well-studied model systems, it is important to maintain a sense of humility and avoid over-generalizing current empirical findings. We thus acknowledge the hubris of attempting to impose conceptual order on a set of systems that have undoubtedly developed \emph{in vivo} from many diverse evolutionary precursors to serve many diverse functional purposes. 


We also caution against overzealous usage of optimization arguments.  
Even if one can successfully identify measures of performance relevant for natural selection, biomolecular machines may only approach, rather than reach, optimal performance due to genetic drift~\cite{lynch10}, finite evolutionary time and path dependence, and shifting situations. 

Finally, we beg pardon for possibly excluding or downplaying exciting new research areas and results; any such short shrift  likely stemmed not from our judgments of importance, but rather from the necessity of concision and our own ignorance.

\begin{acknowledgement}
The authors thank Nancy Forde, Chapin Korosec, John Bechhoefer, Steve Large, Miranda Louwerse, Martin Zuckermann, Joseph Lucero, and Emma Lathouwers (SFU Physics), Jason Wagoner (Laufer Center, Stony Brook), and Hong Qian (Washington Applied Math) for insightful and helpful comments on the manuscript. 
This work was supported by a Natural Sciences and Engineering Research Council of Canada (NSERC) Discovery Grant (DAS) and by a Tier-II Canada Research Chair (DAS).
\end{acknowledgement}


\providecommand{\latin}[1]{#1}
\providecommand*\mcitethebibliography{\thebibliography}
\csname @ifundefined\endcsname{endmcitethebibliography}
{\let\endmcitethebibliography\endthebibliography}{}

\newpage
Aidan Brown is a postdoctoral fellow at the University of California, San Diego. Dr.~Brown received his B.Sc.\ in Physics from the University of Guelph, and his M.Sc.\ and Ph.D.\ in Physics from Dalhousie University. He was a postdoctoral fellow at Simon Fraser University.

David Sivak is Assistant Professor of Physics at Simon Fraser University (SFU), an Associate member of the SFU Departments of Chemistry and Molecular Biology \& Biochemistry, and a Canada Research Chair (Tier II) in Nonequilibrium Statistical Biophysics.
Dr.~Sivak received his A.B.\ in Applied Mathematics from Harvard University, his B.A.\ in Philosophy, Politics, and Economics from Lincoln College, University of Oxford, and his Ph.D.\ in Biophysics from the University of California, Berkeley.
He was a Physicist Postdoctoral Fellow at Lawrence Berkeley National Laboratory and an independent Systems Biology Fellow at the University of California, San Francisco. 
The Sivak group develops nonequilibrium statistical physics and stochastic thermodynamics with applications to biological transduction of free energy and information. 

\end{document}